\documentclass[copyright,creativecommons]{eptcs}
\providecommand{\event}{UITP 2016} 
\usepackage{amsmath}
\usepackage{amsthm}
\usepackage{subcaption}
\usepackage{wrapfig}
\usepackage{tikz}
\usetikzlibrary{calc}
\usepackage{xcolor}
\usepackage{todonotes}
\usepackage{speedith}

\title{Tactical Diagrammatic Reasoning\footnote{This work is supported by EPSRC grant EP/M011763/1.}}
\author{Sven Linker
\institute{Department of Computer Science\\ 
University of Liverpool, UK} 
\email{s.linker@liverpool.ac.uk}
\and Jim Burton
\institute{Visual Modelling Group\\ University of Brighton, UK}
\email{j.burton@brighton.ac.uk}
\and
Mateja Jamnik
\institute{Computer Laboratory\\ University of Cambridge, UK}
\email{Mateja.Jamnik@cl.cam.ac.uk}
}

\def\titlerunning{Tactical Diagrammatic Reasoning}
\def\authorrunning{S. Linker, J. Burton \& M. Jamnik}

\RequirePackage{todonotes}
\presetkeys{todonotes}{fancyline, color=blue!30}{}

\begin{document}

\maketitle

\begin{abstract}
  Although automated reasoning with diagrams has been possible for
  some years, tools for diagrammatic reasoning are generally much less
  sophisticated than their sentential cousins. The tasks of exploring
  levels of automation and abstraction in the construction of proofs
  and of providing explanations of solutions expressed in the proofs
  remain to be addressed. In this paper we take an interactive proof
  assistant for Euler diagrams, Speedith, and add tactics to its
  reasoning engine, providing a level of automation in the
  construction of proofs. By adding tactics to Speedith's repertoire
  of inferences, we ease the interaction between the user and the
  system and capture a higher level explanation of the essence of the
  proof. We analysed the design options for tactics by using metrics
  which relate to human readability, such as the number of inferences
  and the amount of clutter present in diagrams. Thus, in contrast to
  the normal case with sentential tactics, our tactics are designed to
  not only prove the theorem, but also to support explanation.

\end{abstract}

\section{Introduction}

Automated and interactive reasoning with heterogeneous and purely
diagrammatic systems has been possible since the
1990s. Hyperproof~\cite{barwise:h}, 
Openproof~\cite{barker-plummer:oafffhr},
Diabelli~\cite{uj-ijcar2012-diabelli} and 
MixR~\cite{uj-diagrams2014-mixr} are prominent examples of
heterogeneous systems, whilst \textsc{Diamond}~\cite{jamnik:mrwd} and
EDITH~\cite{stapleton:atpieds} demonstrate differing purely
diagrammatic approaches. Speedith~\cite{Urbas2015} is a more
recent interactive theorem prover for spider diagrams, an Euler-based
notation equivalent to monadic first order logic with
equality~\cite{howse:sd}. None of these tools make use of
tactics. Tactical theorem provers enable the strategic application of
sequences of inference rules, providing a form of semi-automated
theorem proving.

As part of an investigation into the readability of diagrammatic
proofs\footnote{\url{http://readableproofs.org}}, we implemented the
first tactical diagrammatic theorem prover by extending
Speedith\footnote{Our extended version of Speedith, alongside a
collection of theorems and proofs, is available from
\url{http://readableproofs.org/speedith}}. Our aims were to provide
users with convenient, high level ways to prove theorems and to enable
tactics to be used as a type of explanation or description of their
proof strategies. The design of our tactics is informed by our
readability investigation, and so we prioritise the presentation and
inspection of steps applied by a tactic. Thus, we make efforts to
ensure that the sequence of inferences involved in the tactic are
coherent for human readers. We do this by sequencing inferential steps
in ways which are, arguably, similar to the ways in which a human might approach
the problem (see Section~\ref{sec:eval}), and which result in diagrams
with favourable properties such as containing less syntax than
diagrams produced by alternative sequences.

In Sections~\ref{sec:background} and~\ref{sec:speedith} we provide context for the work by
explaining the recent history of tools for diagrammatic reasoning. Then in
Section~\ref{sec:tactics} we describe our tactics and their implementation. In
Section~\ref{sec:eval} we evaluate our work by the use of metrics relating to the readability
concerns; these metrics include measures such as the length of the resulting proof, the amount
of ``clutter'' or redundant syntax present in each diagram, and so on. Clutter is relevant
because several empirical studies have shown that it has a negative impact on user
comprehension (see, for example,~\cite{linker:mucoiried}). Finally, we conclude by
describing our goals for the future which include a plan to augment Speedith further with fully
automated reasoning and to use it in educational settings to teach diagrammatic reasoning.



\section{\label{sec:background}Background}

Diagrammatic logics such as Euler diagrams have a number of striking
differences from sentential logics. One of the most important of these
is the fact that diagrams can reveal certain consequences of logical
statements that would require several inferential steps in sentential
systems. This property is known as a ``free
ride''~\cite{shimojima:iaecogrsasg}. In addition, empirical studies
have found that logical diagrams can be more accessible to users without
training~\cite{Mineshima2014}, and thus have the potential to bring
formal reasoning to wider audiences.

\begin{figure}[t]
  \centering
  \includegraphics[width=0.8\textwidth]{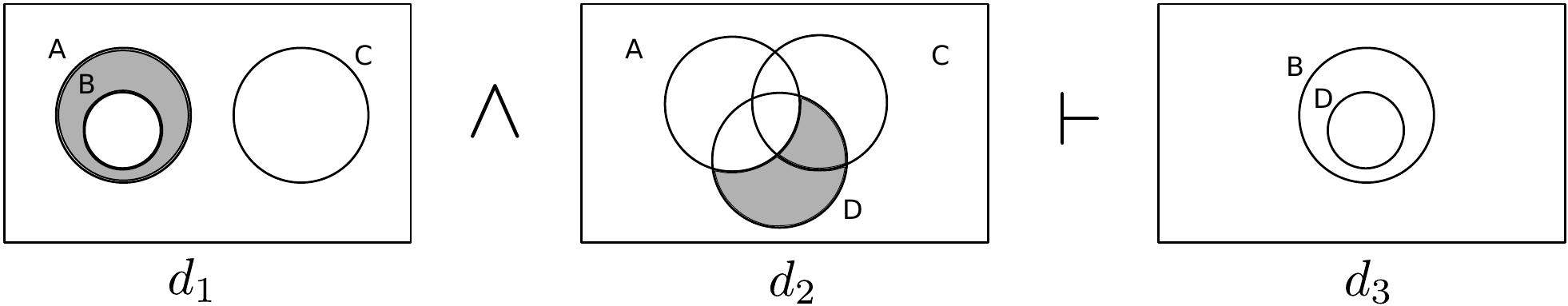}
  \caption{An Euler Theorem}
  \label{fig:euler}
\end{figure}

Our work focuses on reasoning with Euler diagrams, a simple and
well-known formalism in which the properties of a diagram (the
containment, overlap or separateness of circles) directly reflects the
underlying meaning (the subsumption, intersection and disjointness of
sets). This correspondence between the concrete syntax of diagrams and
the semantic level is known as \emph{well
  matchedness}~\cite{gurr:edcsspi} or, in the terminology of the
philosopher C. S. Peirce~\cite[p126]{atkin:p},
\emph{iconicity}. Figure~\ref{fig:euler} shows three unitary diagrams
(that is, ones which do not contain sentential logical
operators). Unitary diagrams can be joined by conjunction and disjunction
to form compound diagrams, and may also be negated by drawing a
horizontal bar above a unitary diagram; in this work we deal only with
unitary and conjunctive compound diagrams. 
In Figure~\ref{fig:euler}, diagram $d_1$ contains
three \emph{contours} labelled $A$, $B$ and $C$ representing
sets. From the placement of the circles we can see that the
\emph{region} $A \cap C$ is not drawn in $d_1$: this means that $A
\cap C = \emptyset$, since missing regions represent the empty set. A
\emph{shaded} region means that the set it represents is empty. For
example, from the shading inside $A$ but outside of $B$ in $d_1$ we
see that the set $A - B$ is empty. Each diagram contains a number of
\emph{zones}, which are determined by the set of contours they are
inside. Thus, there is a zone outside of all contours, one inside $A$
but outside $B$ and $C$, one inside $B$ only, and one inside $C$ only:
therefore, $d_1$ contains 4 zones. A diagram which contains all
possible zones for a given set of contours (that is, all possible
overlaps of the contours are drawn) is said to be in \emph{Venn form}
(for example, diagram \(d_2\) in Figure~\ref{fig:euler}). If a diagram does not contain all
possible zones, the zones that are not drawn are
\emph{missing}. Observe that the emptiness of an intersection can be
denoted by shading a given zone, or by not drawing it at
all. Inference rules for Euler diagrams carry out logically valid
operations such as adding and removing contours and shaded zones, and
combining the information from several diagrams. In
Figure~\ref{fig:euler}, $d_3$ is a consequence of $d_1 \wedge d_2$.

The first automated reasoner for Euler diagrams was
EDITH~\cite{stapleton:atpieds}, which came equipped with different
sound and complete sets of rules and carried out an A* proof search
guided by heuristics~\cite{Hart1968}. EDITH found the shortest proof for a theorem,
given a particular set of rules and provided that a proof exists.

Euler diagrams are the basis for many more expressive notations, such
as spider diagrams~\cite{howse:sd}. Speedith~\cite{Urbas2015} was
implemented as an interactive proof assistant for spider
diagrams. Speedith makes use of iCircles~\cite{stapleton:dedwcttop}, a
Java library which draws Euler diagrams using only circles, and adds
the functionality needed to represent spiders (existential
elements). Although our work is concerned only with Euler diagrams, we
chose Speedith as the basis for extension due to its code base with
effective visualisation of all Euler diagrams.

Sentential tactical reasoning, as found in tools such as
Isabelle~\cite{Paulson1994}, is a powerful semi-automation feature
which applies a specific sequence of inference
rules. It is sometimes designed to reflect the theorem proving
strategies used by humans. Tactics may be used in conjunction with
individual inference steps, for instance, in order to simplify the
problem before addressing individual cases, or used as high level
strategies which may construct the entire proof. The steps taken by a
sentential tactic are not normally displayed to the user, but may be made visible
if required (e.g., Coq uses a command called \texttt{Info} for
this). When those steps are displayed, they may be too low level to
provide any insight to the user, since sentential tactics are not generally
designed with human readers in mind. Since our overall goal is to
design perspicuous diagrammatic proofs, and since clarity is one of
the key motivations for diagrammatic reasoning in the first place, our
tactics are designed with the expectation that users will trace the
results of their application.



\section{\label{sec:speedith}Speedith}

Speedith 
consists of a reasoning kernel,
which can be used to apply rules to the abstract syntax of a given
diagram, and a user interface based on
iCircles 
to draw the diagrams.  In the
following, we will use the term \emph{subgoal} to refer to a compound
diagram within a proof of Speedith. A \emph{proof state} is a list of
subgoals that still have to be solved. Finally, a \emph{proof} in
Speedith is a list of proof states, \(g_0, \dots, g_n\), where for all
\(0 \leq i < n\), there is a subgoal \(d\) in \(g_i\) and \(d^\prime\)
in \(g_{i+1}\) such that \(d^\prime\) is the result of applying one of
the inference rules to \(d\).
 
The rules can be interactively applied to a subgoal. The position of a
subgoal within its proof state is called its \emph{index}.  Speedith
applies rules backwards; the reasoner tries to reduce a
subgoal to the empty diagram, which is a tautology. Hence, we call a
proof, \(p = g_0,\dots, g_n\), \emph{finished} if \(g_n\) is the empty
diagram.

The set of rules within Speedith can be divided into \emph{purely
  logical rules} and \emph{diagrammatic rules}. Rules in the former
set directly correspond to rules and equivalences of propositional
logic, while rules in the latter set manipulate the diagrammatic
structure of one or more unitary diagrams. In this paper, we are only
concerned with the subset of Speedith rules that are applicable to Euler
diagrams.  Furthermore, we only describe the subset of rules we used
in the definition of our tactics. This set is not logically
complete, but extending the application to include a complete set would be technically
straightforward.

\begin{figure}
    \begin{subfigure}{.16\linewidth}
      \begin{center}
         \begin{tikzpicture}
           \node (1) {\includegraphics[height=1.5cm]{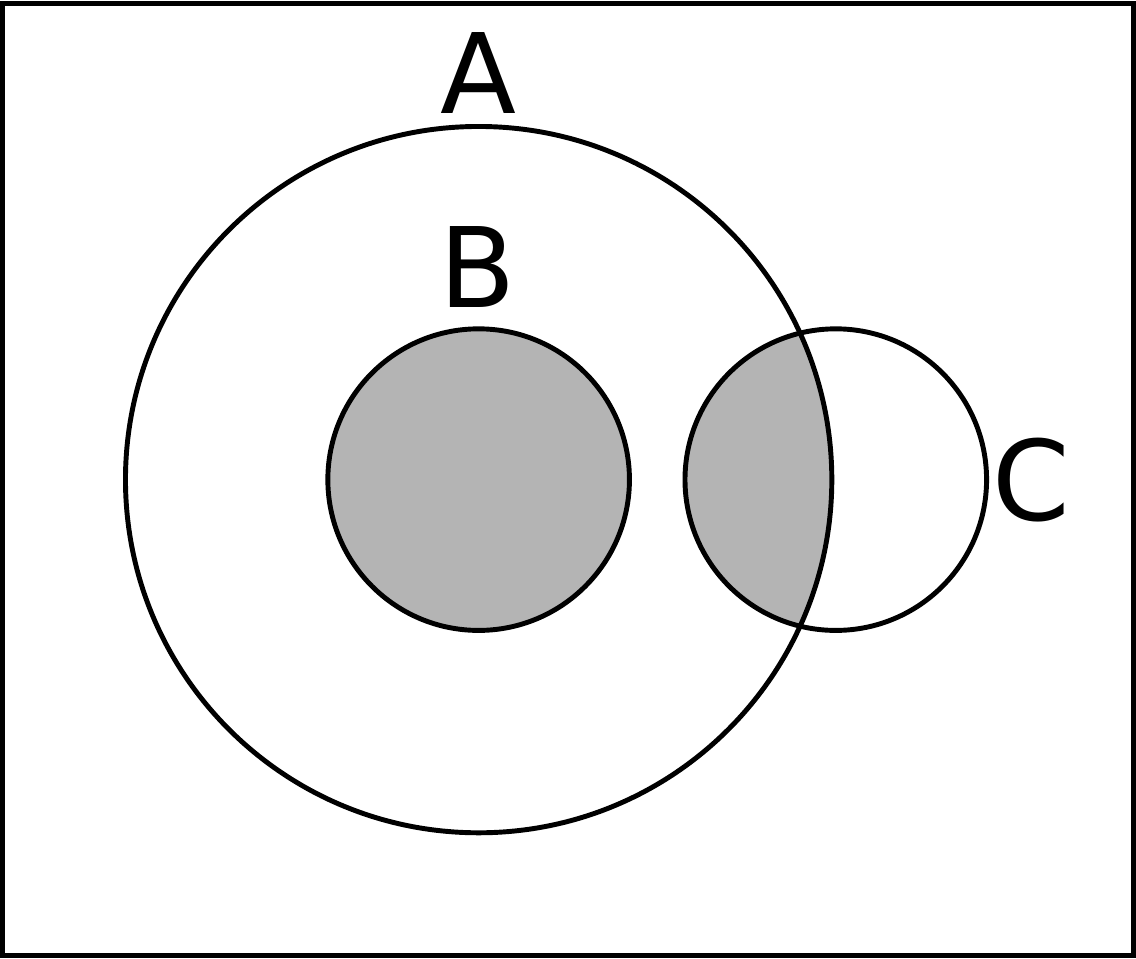}};
           \node[below=of 1, yshift=.75cm] (ecConc) {\includegraphics[height=1.5cm]{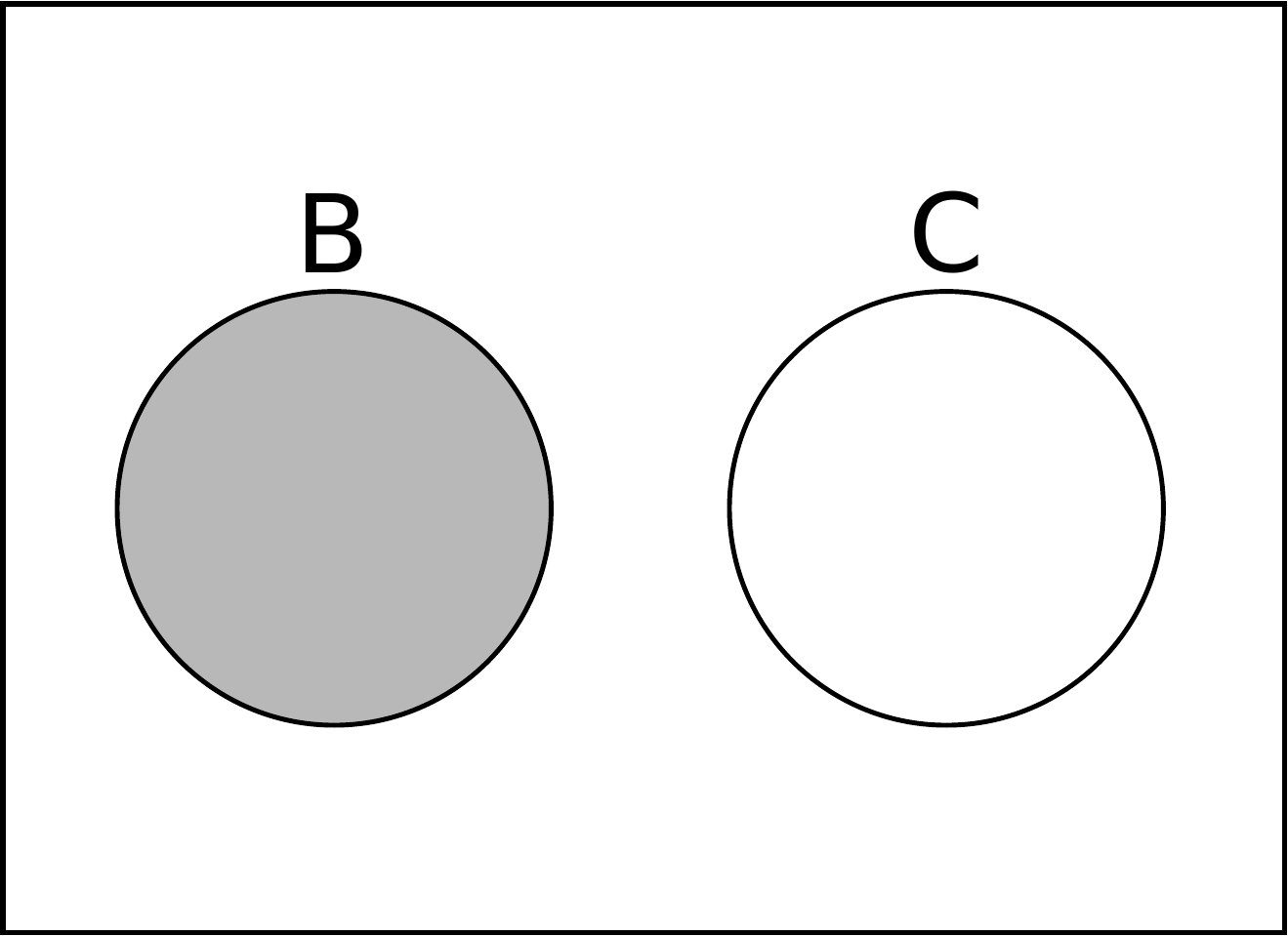}};
           \draw[thick] ($(1.south west) + (0,-.125)$) -- ($(1.south east) + (0,-.125)$);
         \end{tikzpicture}      
      \end{center}
      \caption{Erase Contour}
      \label{fig:ec}
    \end{subfigure}    \begin{subfigure}{.19\linewidth}
      \begin{center}
         \begin{tikzpicture}
           \node (1) {\includegraphics[height=1.5cm]{figures/eraseShadingPrem}};
           \node[below=of 1, yshift=.75cm] (ecConc) {\includegraphics[height=1.5cm]{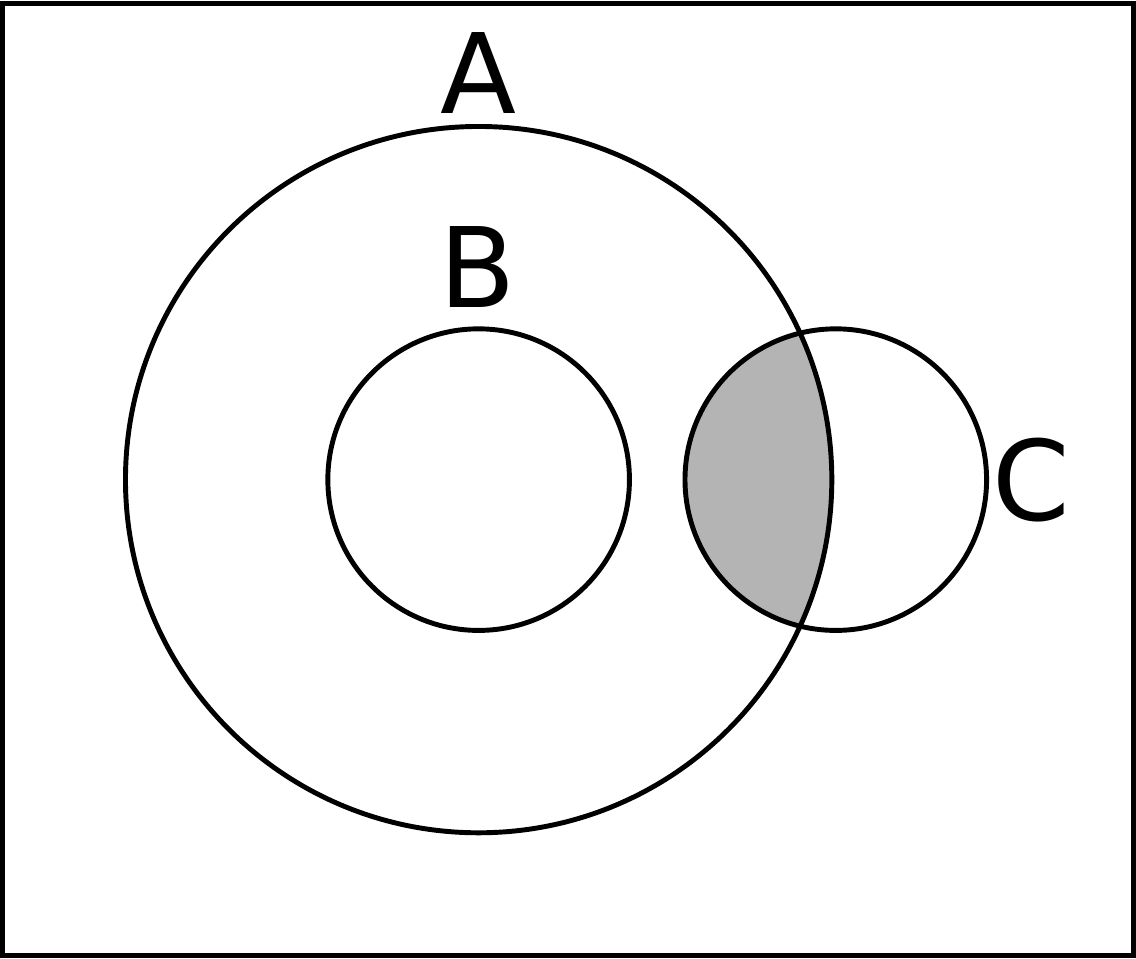}};
           \draw[thick] ($(1.south west) + (0,-.125)$) -- ($(1.south east) + (0,-.125)$);
         \end{tikzpicture}      
      \end{center}
      \caption{Erase Shading}
      \label{fig:es}
    \end{subfigure}
    \begin{subfigure}{.19\linewidth}
      \begin{center}
         \begin{tikzpicture}
           \node (1) {\includegraphics[height=1.5cm]{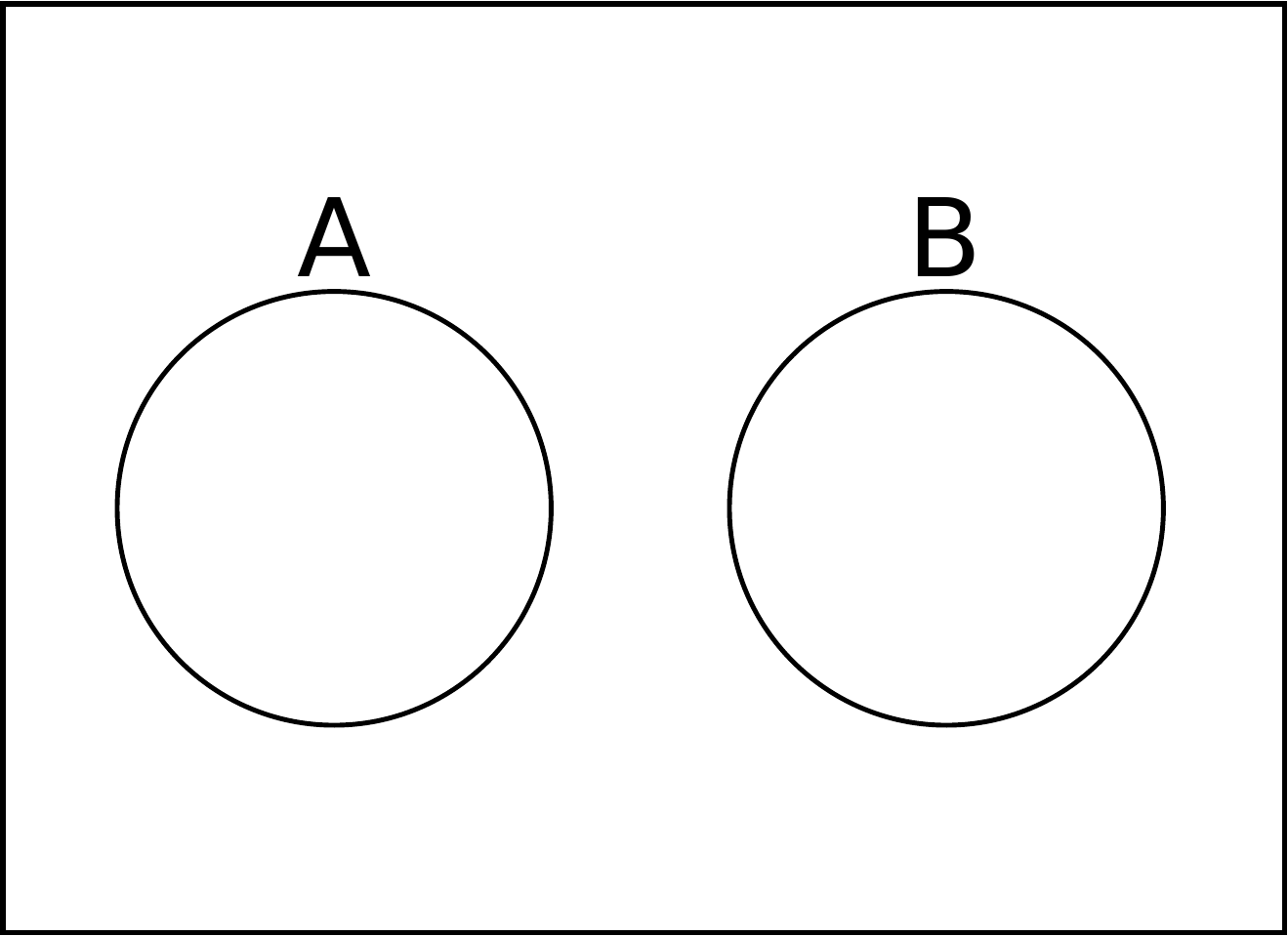}};
           \node[below=of 1, yshift=.75cm] (ecConc) {\includegraphics[height=1.5cm]{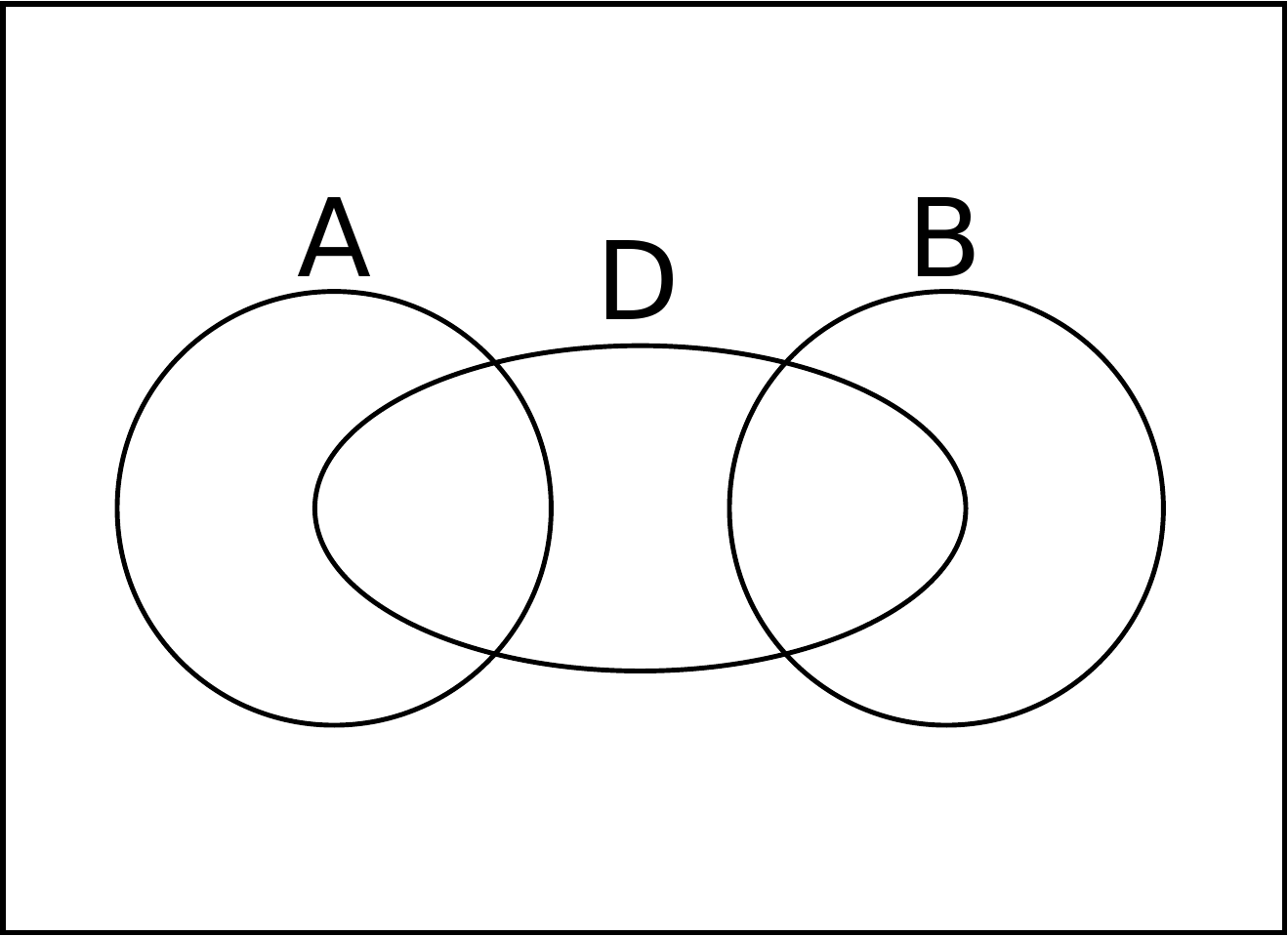}};
           \draw[thick] ($(1.south west) + (0,-.125)$) -- ($(1.south east) + (0,-.125)$);
         \end{tikzpicture}      
      \end{center}
      \caption{Intr. Contour}
      \label{fig:ic}
    \end{subfigure}
    \begin{subfigure}{.2\linewidth}
      \begin{center}
         \begin{tikzpicture}
           \node (1) {\includegraphics[height=1.5cm]{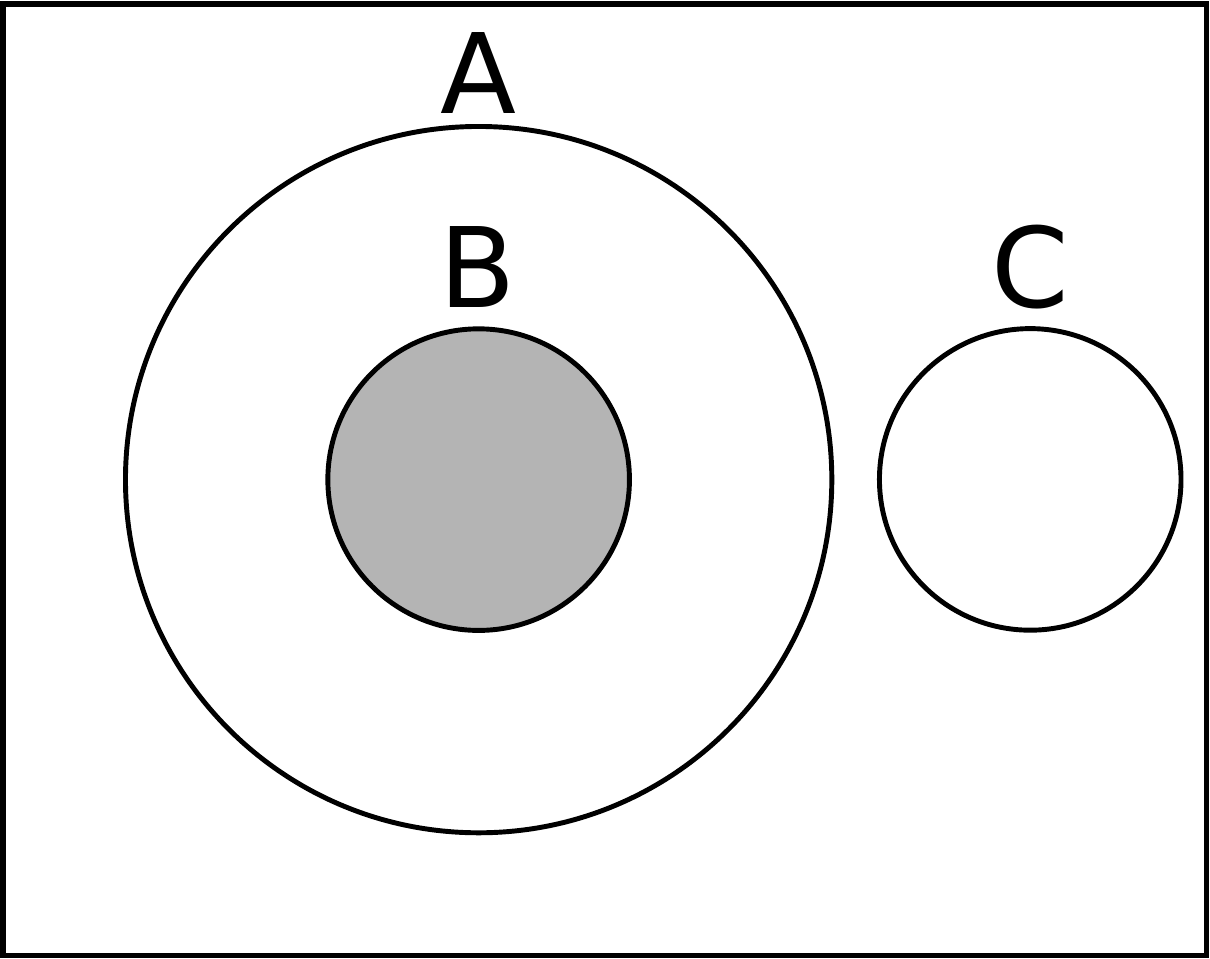}};
           \node[below=of 1, yshift=.75cm] (ecConc) {\includegraphics[height=1.5cm]{figures/eraseShadingPrem}};
           \draw[thick] ($(1.south west) + (0,-.125)$) -- ($(1.south east) + (0,-.125)$);
         \end{tikzpicture}      
      \end{center}
      \caption{Intr. Shaded Zone}
      \label{fig:isz}
    \end{subfigure}
    \begin{subfigure}{.2\linewidth}
      \begin{center}
         \begin{tikzpicture}
           \node (1) {\includegraphics[height=1.5cm]{figures/eraseShadingPrem}};
           \node[below=of 1, yshift=.75cm] (ecConc) {\includegraphics[height=1.5cm]{figures/removeShZoneConc}};
           \draw[thick] ($(1.south west) + (0,-.125)$) -- ($(1.south east) + (0,-.125)$);
         \end{tikzpicture}      
      \end{center}
      \caption{Rem. Shaded Zone}
      \label{fig:rsz}
    \end{subfigure}
    \caption{Examples of Rules~\ref{r:ec}-\ref{r:rsz}}
  \label{fig:ex_ec_es}
\end{figure}

\begin{figure}
    \begin{subfigure}{.3\linewidth}
      \addtocounter{subfigure}{5}
      \begin{center}
         \begin{tikzpicture}
           \node (1) {\includegraphics[height=1.5cm]{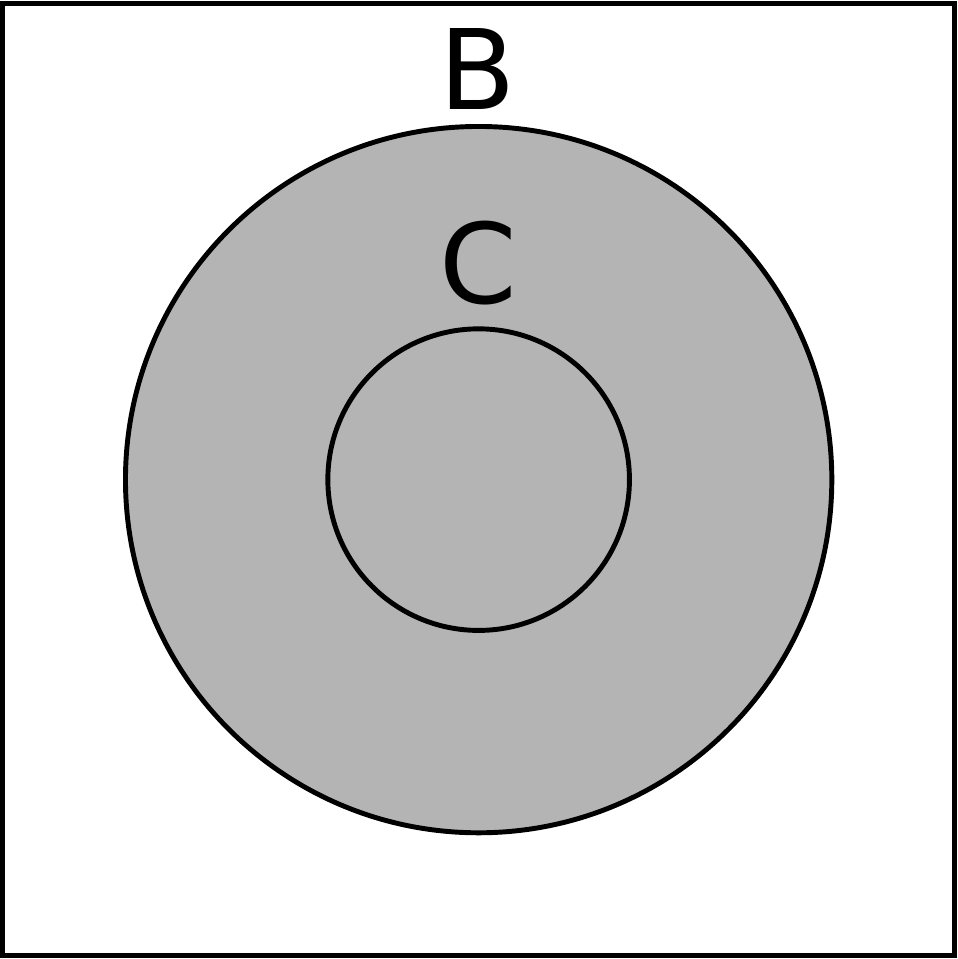}};
           \node[right=of 1, xshift=-.75cm] (2) {\includegraphics[height=1.5cm]{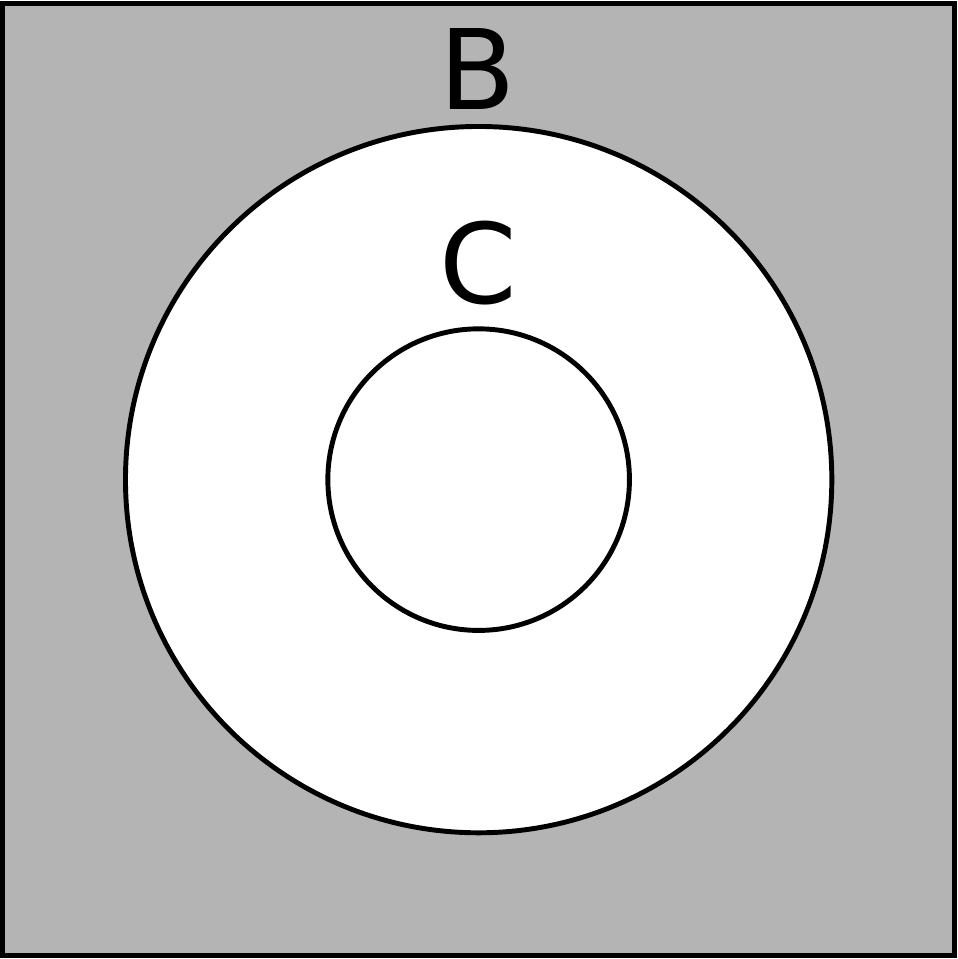}};

           \node[below=of 1, yshift=.75cm, xshift=1cm] (ecConc) {\includegraphics[height=1.5cm]{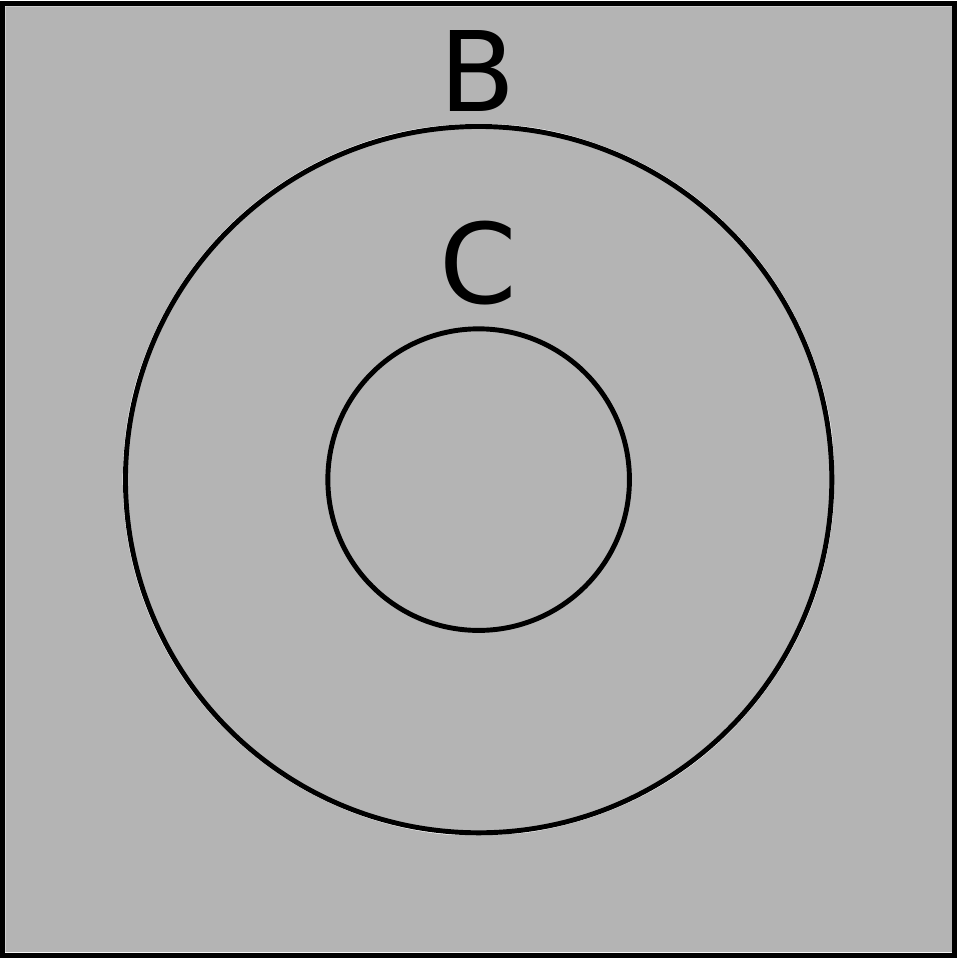}};
           \draw[thick] ($(1.south west) + (0,-.125)$) -- ($(2.south east) + (0,-.125)$);
         \end{tikzpicture}      
      \end{center}
      \caption{Combine}
      \label{fig:co}
    \end{subfigure}
    \begin{subfigure}{.3\linewidth}
      \begin{center}
         \begin{tikzpicture}
           \node (1) {\includegraphics[height=1.5cm]{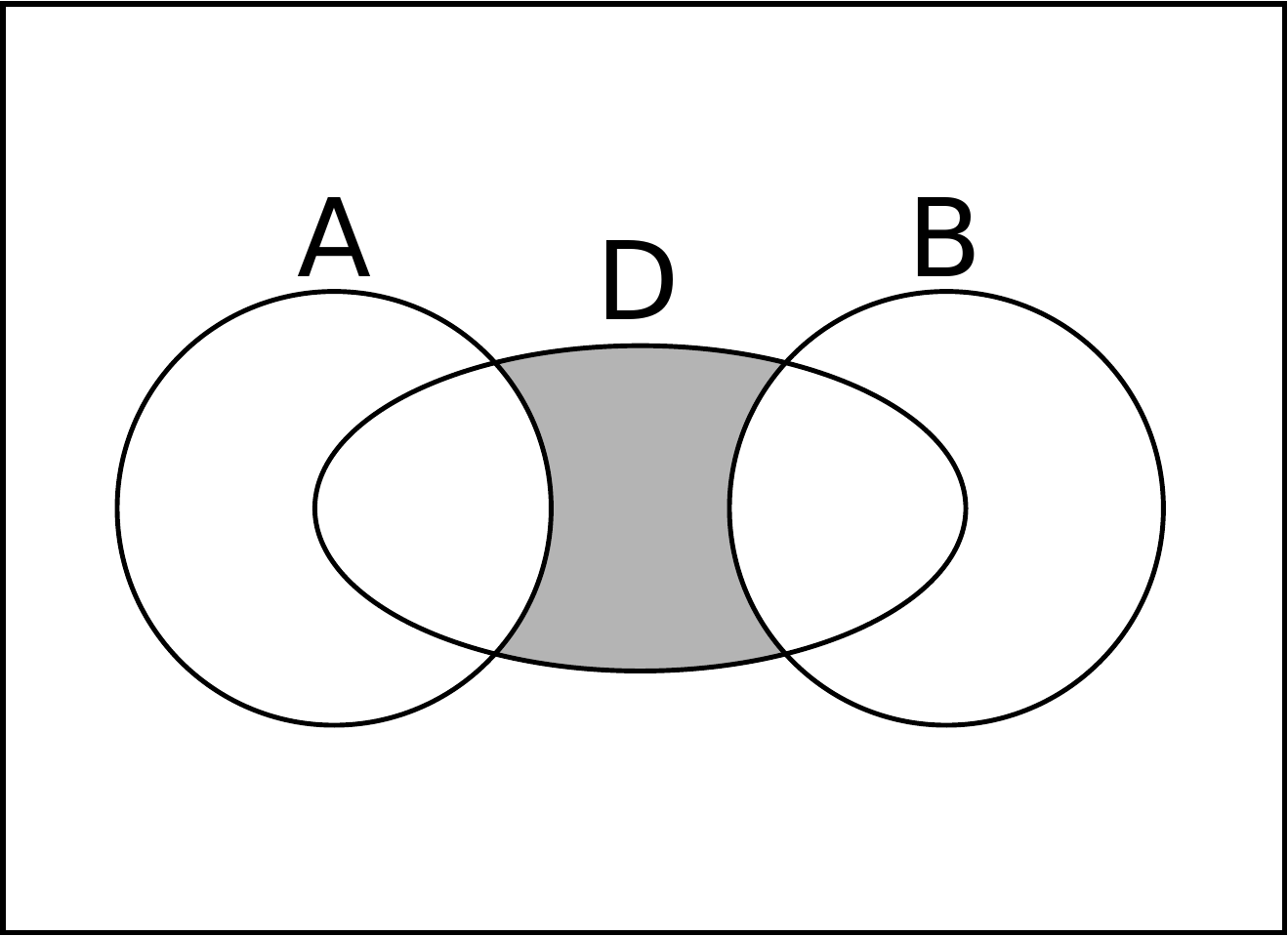}};
           \node[right=of 1, xshift=-.75cm] (2) {\includegraphics[height=1.5cm]{figures/prem2}};

           \node[below=of 1, yshift=.75cm] (ecConc) {\includegraphics[height=1.5cm]{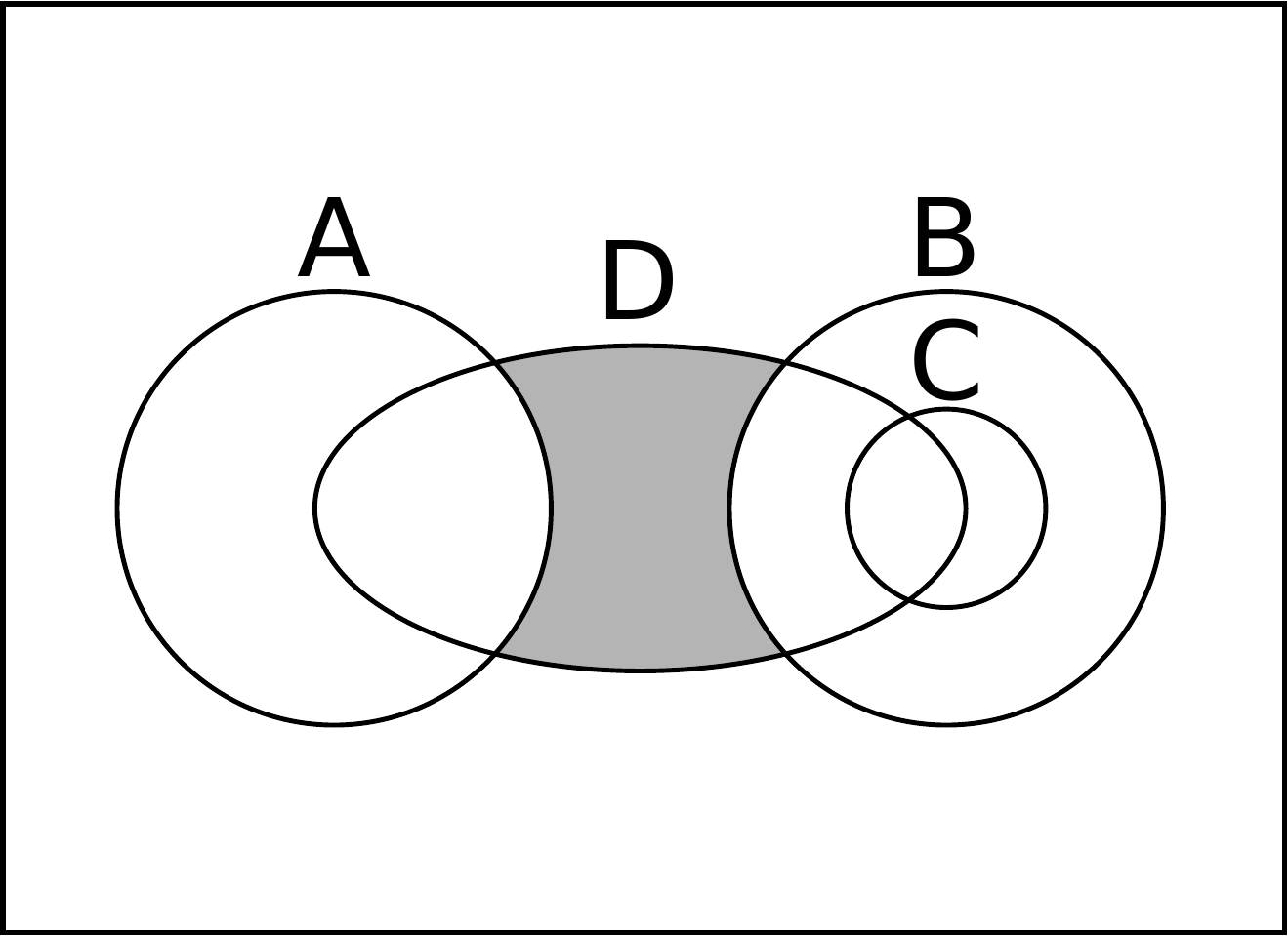}};
           \node[below=of 2, yshift=.75cm] (ecConc2) {\includegraphics[height=1.5cm]{figures/prem2}};
           \draw[thick] ($(1.south west) + (0,-.125)$) -- ($(2.south east) + (0,-.125)$);
         \end{tikzpicture}      
      \end{center}
      \caption{Copy Contour}
      \label{fig:cc}
    \end{subfigure}
    \begin{subfigure}{.3\linewidth}
      \begin{center}
         \begin{tikzpicture}
           \node (1) {\includegraphics[height=1.5cm]{figures/prem1}};
           \node[right=of 1, xshift=-.75cm] (2) {\includegraphics[height=1.5cm]{figures/prem2}};

           \node[below=of 1, yshift=.75cm] (ecConc) {\includegraphics[height=1.5cm]{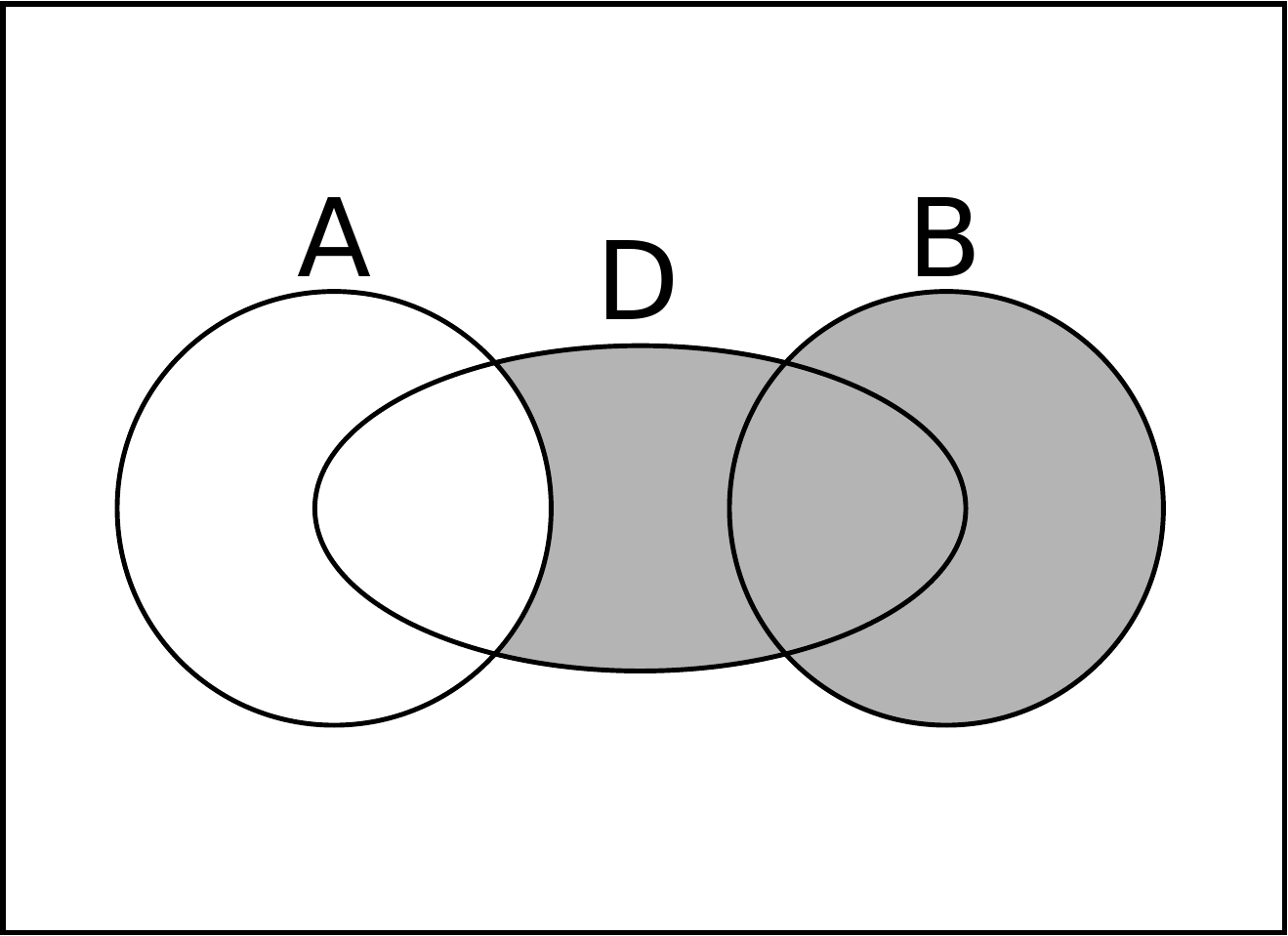}};
           \node[below=of 2, yshift=.75cm] (ecConc2) {\includegraphics[height=1.5cm]{figures/prem2}};
           \draw[thick] ($(1.south west) + (0,-.125)$) -- ($(2.south east) + (0,-.125)$);
         \end{tikzpicture}      
      \end{center}
      \caption{Copy Shading}
      \label{fig:cs}
    \end{subfigure}
    \caption{Examples of Rules~\ref{r:co}-\ref{r:cs}}
    \label{fig:ex_co_copy}
  \end{figure}

We consider the following diagrammatic inference rules:
\begin{enumerate}\setlength{\itemsep}{-1mm}
\item Erase contour, \label{r:ec}
\item Erase shading, \label{r:es}
\item Introduce contour,\label{r:ic}
\item Introduce shaded zone, \label{r:isz}
\item Remove shaded zone, \label{r:rsz}
\item Combine diagrams, \label{r:co}
\item Copy contour, and\label{r:cc}
\item Copy shading. \label{r:cs}
\end{enumerate}
Rules~\ref{r:ec}-\ref{r:rsz} (see Figure~\ref{fig:ex_ec_es}) only
affect single unitary diagrams, while rules~\ref{r:co}-\ref{r:cs} (see
Figure~\ref{fig:ex_co_copy}) take several unitary diagrams within the
premises into account. Furthermore, rules~\ref{r:ec} and \ref{r:es}
remove information from the manipulated diagrams, while
rules~\ref{r:ic}-\ref{r:rsz} are equivalences.

\emph{Erase Contour} (see Figure~\ref{fig:ex_ec_es}.\subref{fig:ec}) removes a contour from
a unitary diagram.  If this contour was separating a shaded zone from
a non-shaded zone, then the unified zone in the result will not be
shaded. This rule removes information, and so the premiss implies the
conclusion but not vice versa. For example, in the premiss of
Figure~\ref{fig:ec}, the contour \(A\) separated the two zones within
contour \(C\). Since one of them is shaded and the other is not,
removing \(A\) also removes the shading within \(C\).

\emph{Erase Shading} (see Figure~\ref{fig:ex_ec_es}.\subref{fig:es}) removes the shading of
a single zone from a unitary diagram. This rule also is not an
equivalence rule since it removes information.

\renewcommand{\thesubfigure}{\arabic{subfigure}}

\emph{Introduce Contour} (see Figure~\ref{fig:ex_ec_es}.\subref{fig:ic}) is used to add a
new contour to a unitary diagram, such that it intersects all visible
zones. In this way, the addition of the new contour does not introduce
new information into the unitary diagram. This rule is an equivalence
rule.

Rules~\ref{r:isz} (see Figure~\ref{fig:ex_ec_es}.\subref{fig:isz}) and~\ref{r:rsz} (see
Figure~\ref{fig:ex_ec_es}.\subref{fig:rsz}) are used to replace a missing zone with a shaded
zone within a unitary diagram and vice versa. Both of these rules
preserve equivalence.

When using rule~\ref{r:co} (see Figure~\ref{fig:ex_co_copy}.\subref{fig:co}) to \emph{combine}
two unitary diagrams (also an equivalence), both diagrams must contain
the same set of zones. In the result, the two diagrams are replaced by
a single diagram with the same set of zones, and in which a zone is
shaded if and only if it is shaded in one of the original diagrams.

For the copy rules,~\ref{r:cc} and~\ref{r:cs}, we have to identify
which zones in different diagrams \emph{correspond}~\cite{Howse2002}
to each other; that is, which represent the same underlying
set. \emph{Copy Contour} (see Figure~\ref{fig:ex_co_copy}.\subref{fig:cc}) can be used to copy
a contour, \(c\), from a unitary diagram \(d_1\) to a second unitary
diagram, \(d_2\). The placement of \(c\) respects the topological
relations specified in \(d_1\) between itself and the contours that
occur in both diagrams.  The next rule in our system is \emph{Copy
  Shading} (see Figure~\ref{fig:ex_co_copy}.\subref{fig:cs}). If a set of zones, \(z_1\), is not
shaded in a diagram, \(d_1\), and corresponds to a set of shaded zones, \(z_2\), in
a second diagram, then \emph{Copy Shading} can be used to shade all zones
within \(z_1\). Consider Figure~\ref{fig:ex_co_copy}.\subref{fig:cs}. The two zones within
contour \(B\) in the left premiss and the two zones within \(B\) in
the right premiss correspond to each other (both denote the whole of
\(B\)). Hence we can apply \emph{Copy Shading} to shade two zones
in the left diagram. Both of these rules are equivalence rules.

\renewcommand{\thesubfigure}{\alph{subfigure}}


The only purely logical rule we consider for tactical reasoning is \emph{idempotency}.  It may
be used to remove one of two identical conjuncts. So, from \(d \land d\) we get simply \(d\).



\section{\label{sec:tactics}Tactics}

Here, we give an overview of the implemented tactics and their
restrictions.  The implementation of tactics was originally inspired
by the Isabelle 
interactive symbolic theorem
prover, but differs in a number of ways: the choice of programming
language is Scala~\cite{Odersky2004}, the features are distinctive to
diagrammatic reasoning, and our goal is for users to be able to inspect
the proof. In sentential reasoners, the user is typically shown only
the remaining subgoals after applying a tactic. In contrast, we
include a feature within Speedith to display a full proof including
all inference rule applications.

Each tactic is of the type \(\typeGoals \to \typeInt \to \typeResult \to \typeOption[\typeResult]\).
 It takes as parameters the current proof state (of the type \(\typeGoals\)) and the index
of the subgoal to which the tactic will be
 applied. The final parameter, of the type \(\typeResult\), contains
the list of rules that were applied by tactics during the current
invocation of a tactic by the user and the proof state after applying these rules. This result will only be updated
by rule-level tactics (see Section~\ref{sec:rule-level-tactics}).
The type \(\typeOption[A]\) in Scala
corresponds to the type of the same name in ML~\cite{Milner1997}. It consists of two constructors \(\mathit{Some}\)
and \(\mathit{None}\), where the first takes a parameter of type \(A\). Since the application
of a tactic may fail, it can either return a new \(\typeResult\) (the updated list of
rule applications) or nothing.
The tactics can only be applied to a subgoal of the form \(D \to D^\prime\),
where the consequent \(D^\prime\) is a single unitary diagram and the antecedent \(D\) is a conjunction of
several unitary diagrams. 
This is a deliberate restriction for simplicity.


To combine tactics and build complex tactics from simpler ones,
we implemented combinators or \emph{tacticals} in the fashion of
Isabelle: see Table~\ref{tab:combinators}, where the second column
shows the functional type of each tactical.

\begin{table}[t]
  \centering
\caption{Tacticals of Speedith}
\label{tab:combinators}
\begin{tabular}[htb]{|l|l|}
\hline
Name & Type \\
\hline
\emph{THEN}         & \(\typeTactic \to \typeTactic \to \typeTactic\)\\
\emph{ORELSE}       & \(\typeTactic \to \typeTactic \to \typeTactic\)\\
\emph{REPEAT}       & \(\typeTactic \to \typeTactic \)\\
\emph{COND}         & \(\typeGoalPredicate \to \typeTactic \to \typeTactic \to \typeTactic\)\\
\emph{DEPTH\_FIRST} & \(\typeGoalPredicate \to \typeTactic \to \typeTactic  \)\\
\emph{id}           & \(\typeTactic\)\\
\emph{fail}         & \(\typeTactic\)\\
\hline
\end{tabular}
\end{table}

The combinator \emph{THEN} takes two tactics as parameters, executes
the first, and subsequently executes the second. If either tactic
fails, so does the tactic formed by their combination. \emph{ORELSE},
on the other hand, executes the second tactic only if the first tactic
fails. Hence the combination only fails if both tactics fail to
execute on the given subgoal. A \(\typeGoalPredicate\) is a
function from the proof state (a list of subgoals) and the index
of a subgoal to \(\typeBool\).  In the tactic
\emph{COND} the argument of type \(\typeGoalPredicate\) is used to
analyse the current proof state and choose whether the first or the second
argument should be executed. For repeating tactics we employ two
tacticals.  \emph{REPEAT} applies the given tactic until it fails. For
a more guided search, \emph{DEPTH\_FIRST} should be used. It repeats a
tactic until the predicate given as a first parameter returns
true. The tactics \emph{id} and \emph{fail} always succeed or fail,
respectively.

The tactics do not increase the expressiveness of Speedith. For the sake of this description, 
we classify the tactics we implemented into three categories of increasing complexity: \emph{rule-level},
\emph{low-level} and \emph{high-level}. 
Rule-level tactics apply a single rule once.
Low-level tactics combine rule-level tactics to manipulate the
premises in a certain way. For instance, Tactic~\ref{tac:iamz},
described in Section~\ref{sec:low-level-tactics} below, converts the
unitary diagrams in the antecedent into Venn diagrams, that is,
diagrams without missing zones. Finally, high-level tactics may use
both rule-level and low-level tactics to implement a particular
reasoning strategy.

Currently, users can choose a tactic to be applied from a menu,
similar to the way single rules are chosen. By default, only the
high-level tactics are shown in the menu, but users can change the
preferences to show also the low-level tactics. Since rule-level
tactics are simply applications of single rules, they do not
appear in the tactics menu.

\subsection{Rule-Level Tactics}
\label{sec:rule-level-tactics}

There is no notion of higher-order resolution for diagrams. Thus, we implemented a tactic for
each individual rule, that is, each rule-level tactic applies exactly one of the rules. In that
way, soundness of the tactics is dependent on the soundness of the implementation of the set of
rules. Most of the rules need additional arguments. For example, \emph{Erase Contour} must be
supplied with the unitary diagram within the antecedent to which it will be applied and the
name of the contour to be erased.  To achieve this, we employ two types of functions:
\emph{diagram predicates} and \emph{choosers} (see Table~\ref{tab:auxiliary}). The predicate
returns \emph{True} if the accompanying tactic can be applied to a given diagram. The chooser
function is applied to the diagram identified by the predicate to identify the diagrammatic
element required by the tactic. For example, a suitable predicate for the tactic which applies
\emph{Erase Contour} could return \emph{True} if the given diagram is unitary and contains a
contour. A suitable chooser function would return an element from the set of contours;
depending on the context, this might be an arbitrary contour or the chooser function could use
a more complex implementation to choose a candidate that meets certain requirements.  The \(\typeOption\) datatype allows
us to indicate that a chooser function found a suitable element \(a\) of type \(A\) by using
\(\mathit{Some(a)}\) as the return value, or that no such element exists by using
\(\mathit{None}\). Internally, a rule-level tactic traverses the syntax tree of the subgoal
identified by the supplied index and proof state until the diagram predicate evaluates to
\emph{True}, yielding the diagram \(d\).  Then it applies the given chooser function to \(d\)
to get the final parameter. With this information, it can apply the rule it implements and
update the result of the tactic accordingly. If any of these steps fail (i.e., the diagram
predicate does not evaluate to \emph{True} on any subdiagram; the chooser function does not
find a suitable element and returns \(\mathit{None}\); or the application of the rule fails),
the tactic returns \(\mathit{None}\). Otherwise, the return value is the updated list of rule
applications, including the new proof state.
%
\begin{table}
\centering
\caption{Auxiliary Functions}\label{tab:auxiliary}
\begin{tabular}[htb]{|l|l|}
\hline
Name & Type \\
\hline
\emph{DiagramPredicate} & \(\typeDiagram \to \typeBool\)\\
\emph{Chooser}          & \(\typeDiagram \to \typeOption[A]\)\\
\emph{GoalPredicate}    & \(\typeGoals \to \typeInt \to \typeBool\)\\
\hline
\end{tabular}
\end{table}

\subsection{Low-Level Tactics}
\label{sec:low-level-tactics}

Low-level tactics generally try to apply a single rule as often
as possible. In some cases, however, we use additional rules to 
increase the effectiveness of the tactic, where increased effectiveness could mean less clutter
in the resulting diagrams, for instance. The tactics in
this section are essentially stepping stones towards the high-level
tactics in Section~\ref{sec:high-level-tactics}. Recall that all subgoals
are of the form \(D \imp D^\prime\), where the antecedent \(D\) consists of a
conjunction of several unitary diagrams, and the consequent \(D^\prime\) is unitary. All tactics will
try to remove the subgoal whenever it consists of a trivial
implication, that is, a subgoal of the form \(A \imp A\).

\begin{figure}[htb]
  \centering
  \begin{tikzpicture}
   \node (init) {  
     \begin{tikzpicture}
       \node[draw,rectangle](left_conj) {
         \begin{tikzpicture}
           \node (1) {\includegraphics[height=1.5cm]{figures/prem1}};
           \node[right=of 1, xshift=-1.2cm] (and) {\(\land\)};
           \node[right=of and, xshift=-1.2cm] (2) {\includegraphics[height=1.5cm]{figures/prem2}};
         \end{tikzpicture}
       };
       \node[right=of left_conj, xshift=-1.05cm] (outer_and) {\(\land\)};
       \node[right=of outer_and, xshift=-1.05cm](right_conj) {\includegraphics[height=1.5cm]{figures/prem3}};
     \end{tikzpicture}
   };
   \node[above right  =of init, yshift=-2.15cm] (tac1) {
     \begin{tikzpicture}
       \node[draw,rectangle](left_conj) {
         \begin{tikzpicture}
           \node (1) {\includegraphics[height=1.5cm]{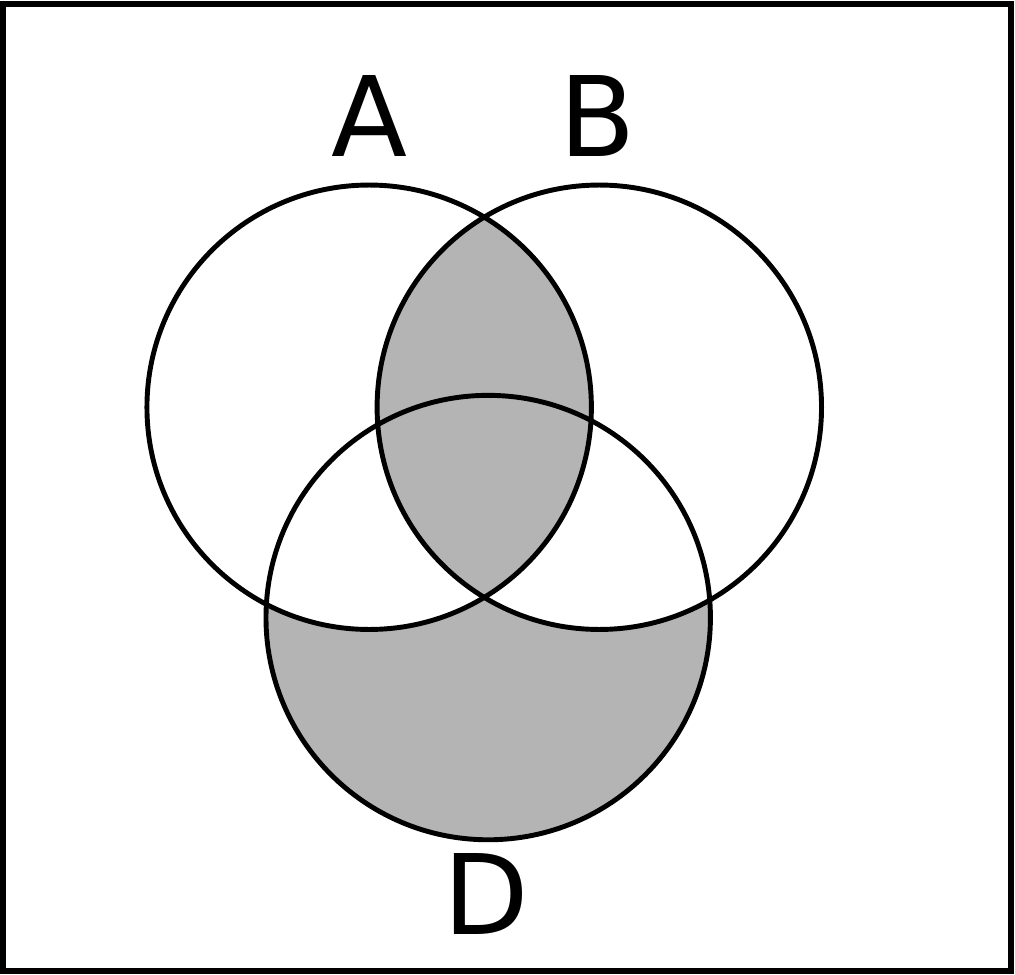}};
           \node[right=of 1, xshift=-1.2cm] (and) {\(\land\)};
           \node[right=of and, xshift=-1.2cm] (2) {\includegraphics[height=1.5cm]{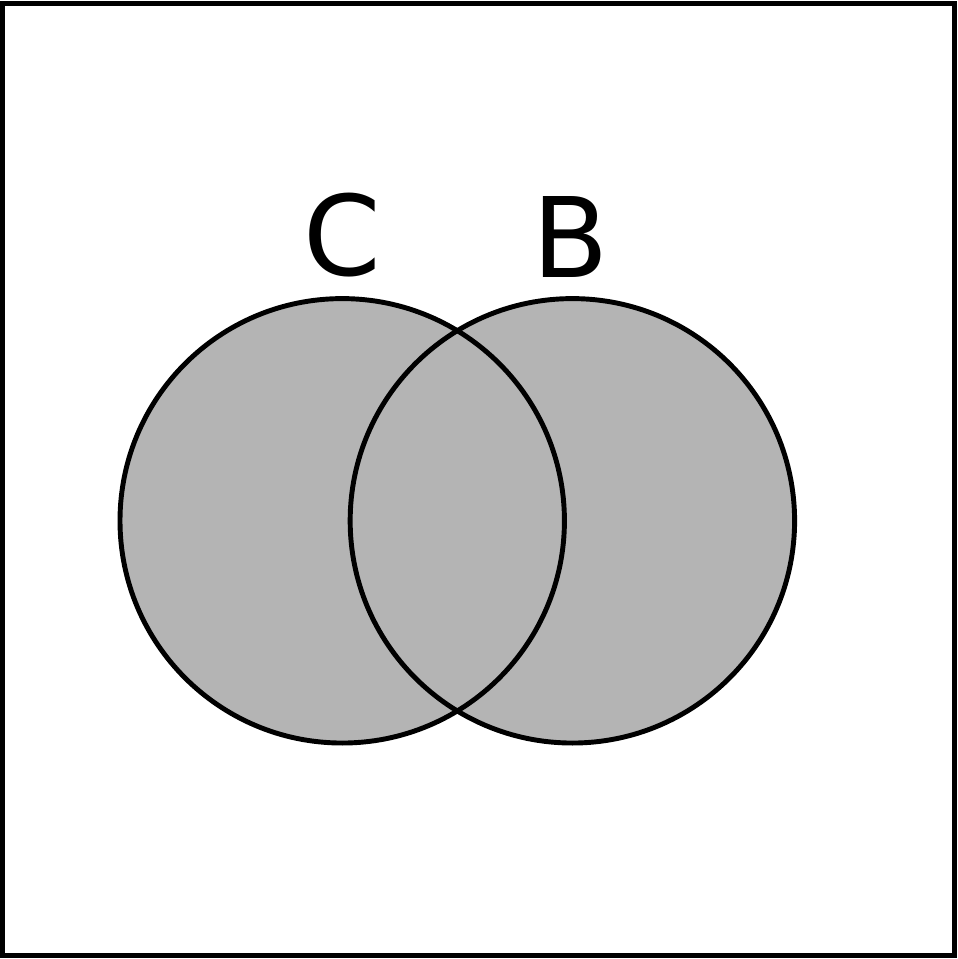}};
         \end{tikzpicture}
       };
       \node[right=of left_conj, xshift=-1.05cm] (outer_and) {\(\land\)};
       \node[right=of outer_and, xshift=-1.05cm](right_conj) {\includegraphics[height=1.5cm]{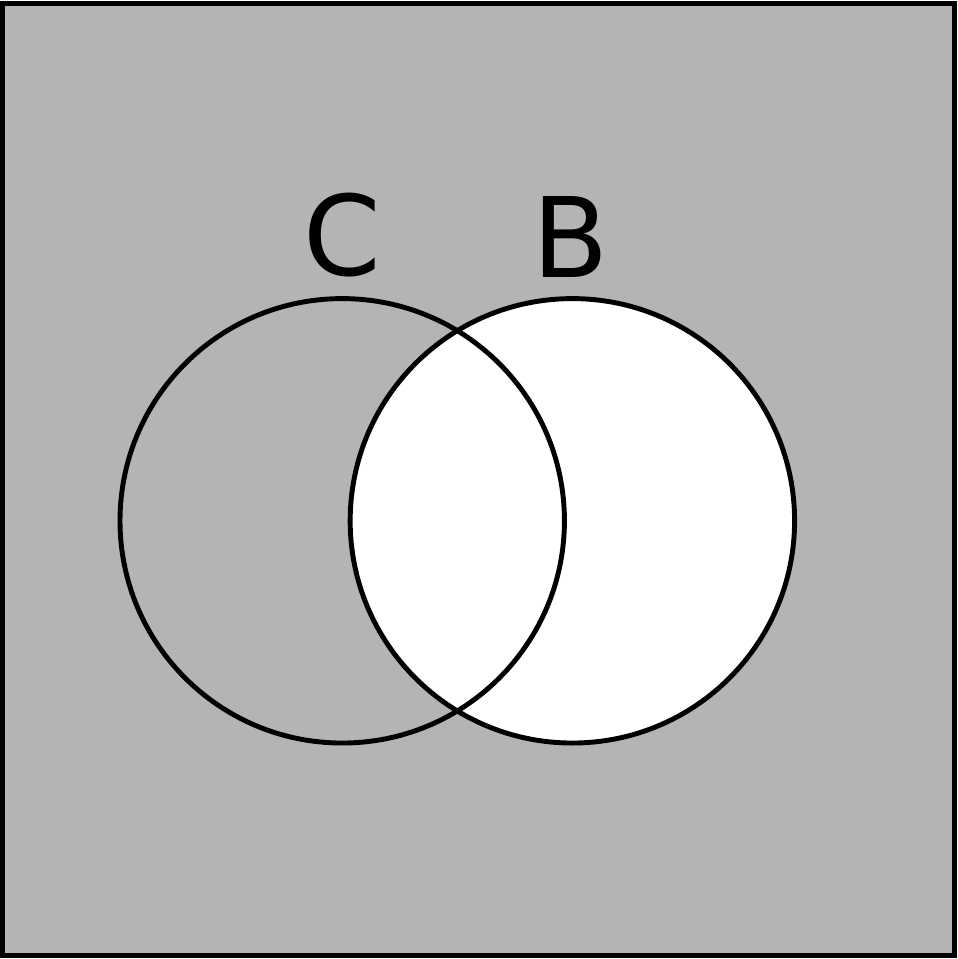}};
     \end{tikzpicture}
   };
   \node[below right =of init, yshift=2.15cm] (tac2) {
     \begin{tikzpicture}
       \node[draw,rectangle](left_conj) {
         \begin{tikzpicture}
           \node (1) {\includegraphics[height=1.5cm]{figures/venn1}};
           \node[right=of 1, xshift=-1.2cm] (and) {\(\land\)};
           \node[right=of and, xshift=-1.2cm] (2) {\includegraphics[height=1.5cm]{figures/venn2}};
         \end{tikzpicture}
       };
       \node[right=of left_conj, xshift=-1.05cm] (outer_and) {\(\land\)};
       \node[ right=of outer_and, xshift=-1.05cm](right_conj) {\includegraphics[height=1.5cm]{figures/prem3}};
     \end{tikzpicture}
   };
   \draw[->](init) to node[above] {Tac.~\ref{tac:iamz}} (tac1);
   \draw[->](init) to node[below] {Tac.~\ref{tac:iamzd}} (tac2);
 \end{tikzpicture}    
 \caption{Examples for Introduce All Shaded Zones}
 \label{fig:ex_intr_sh_zones}
\end{figure}

The following tactics will be used by higher-level tactics to implement ``Venn-style'' reasoning.
This proceeds in three stages: the diagrams in the antecedent are transformed into Venn-form by introducing
all missing zones; contours are introduced to equalise the
contours present in each diagram; 
and finally, the diagrams in the antecedent are iteratively combined into a single
unitary diagram. We created two versions of the introduction
tactics to implement two versions of Venn-style reasoning: breadth-first
and depth-first (cf. Section~\ref{sec:high-level-tactics}). The effect
of these tactics is illustrated in Figure~\ref{fig:ex_intr_sh_zones}. In this figure
and in all of the following examples in this section, we omit the consequent
since the tactics only affect the antecedent.
 
\begin{tac}[Introduce All Shaded Zones]
\label{tac:iamz}
This tactic (see Figure~\ref{fig:ex_intr_sh_zones}) introduces all
missing zones in all unitary diagrams in the antecedent.  Hence, it
transforms the diagrams into Venn form. It uses rule \ref{r:isz}.
\end{tac}

\begin{tac}[Introduce All Shaded Zones (Deepest)]
\label{tac:iamzd}
This tactic (see Figure~\ref{fig:ex_intr_sh_zones}) is similar to
tactic \ref{tac:iamz}, but only introduces shaded zones in a
conjunction of two unitary diagrams. That is, the tactic ``drills down'' in the structure of
the compound diagram to find and apply itself to a pair of unitary diagrams in conjunction.
\end{tac}

\begin{figure}[htb]
  \centering
  \begin{tikzpicture}
   \node (init) {  
     \begin{tikzpicture}
       \node[draw,rectangle](left_conj) {
         \begin{tikzpicture}
           \node (1) {\includegraphics[height=1.5cm]{figures/prem1}};
           \node[right=of 1, xshift=-1.2cm] (and) {\(\land\)};
           \node[right=of and, xshift=-1.2cm] (2) {\includegraphics[height=1.5cm]{figures/prem2}};
         \end{tikzpicture}
       };
       \node[right=of left_conj, xshift=-1.05cm] (outer_and) {\(\land\)};
       \node[ right=of outer_and, xshift=-1.05cm](right_conj) {\includegraphics[height=1.5cm]{figures/prem3}};
     \end{tikzpicture}
   };
   \node[above right  =of init, yshift=-2.15cm] (tac1) {
     \begin{tikzpicture}
       \node[draw,rectangle](left_conj) {
         \begin{tikzpicture}
           \node (1) {\includegraphics[height=1.5cm]{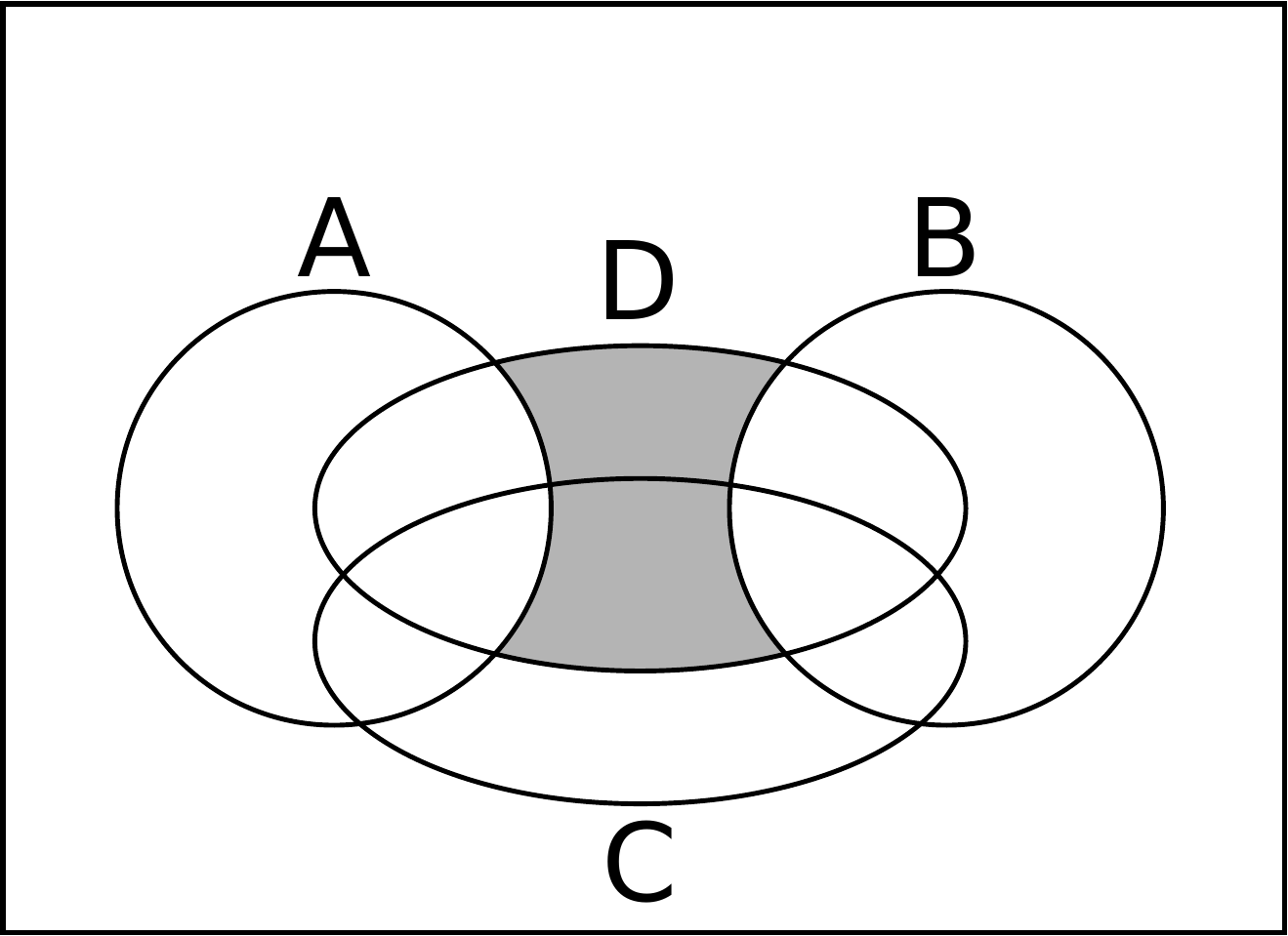}};
           \node[right=of 1, xshift=-1.2cm] (and) {\(\land\)};
           \node[right=of and, xshift=-1.2cm] (2) {\includegraphics[height=1.5cm]{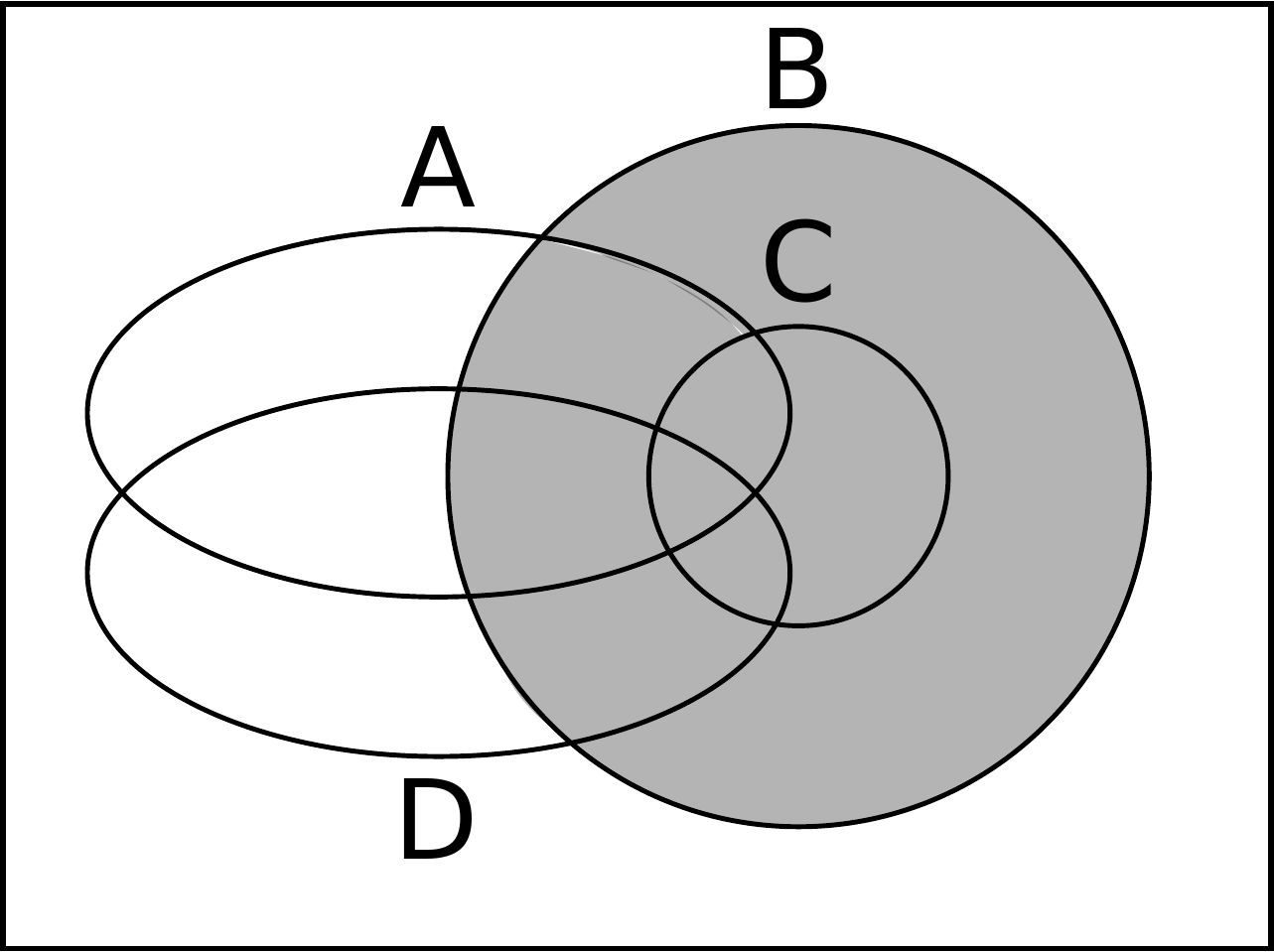}};
         \end{tikzpicture}
       };
       \node[right=of left_conj, xshift=-1.05cm] (outer_and) {\(\land\)};
       \node[ right=of outer_and, xshift=-1.05cm](right_conj) {\includegraphics[height=1.5cm]{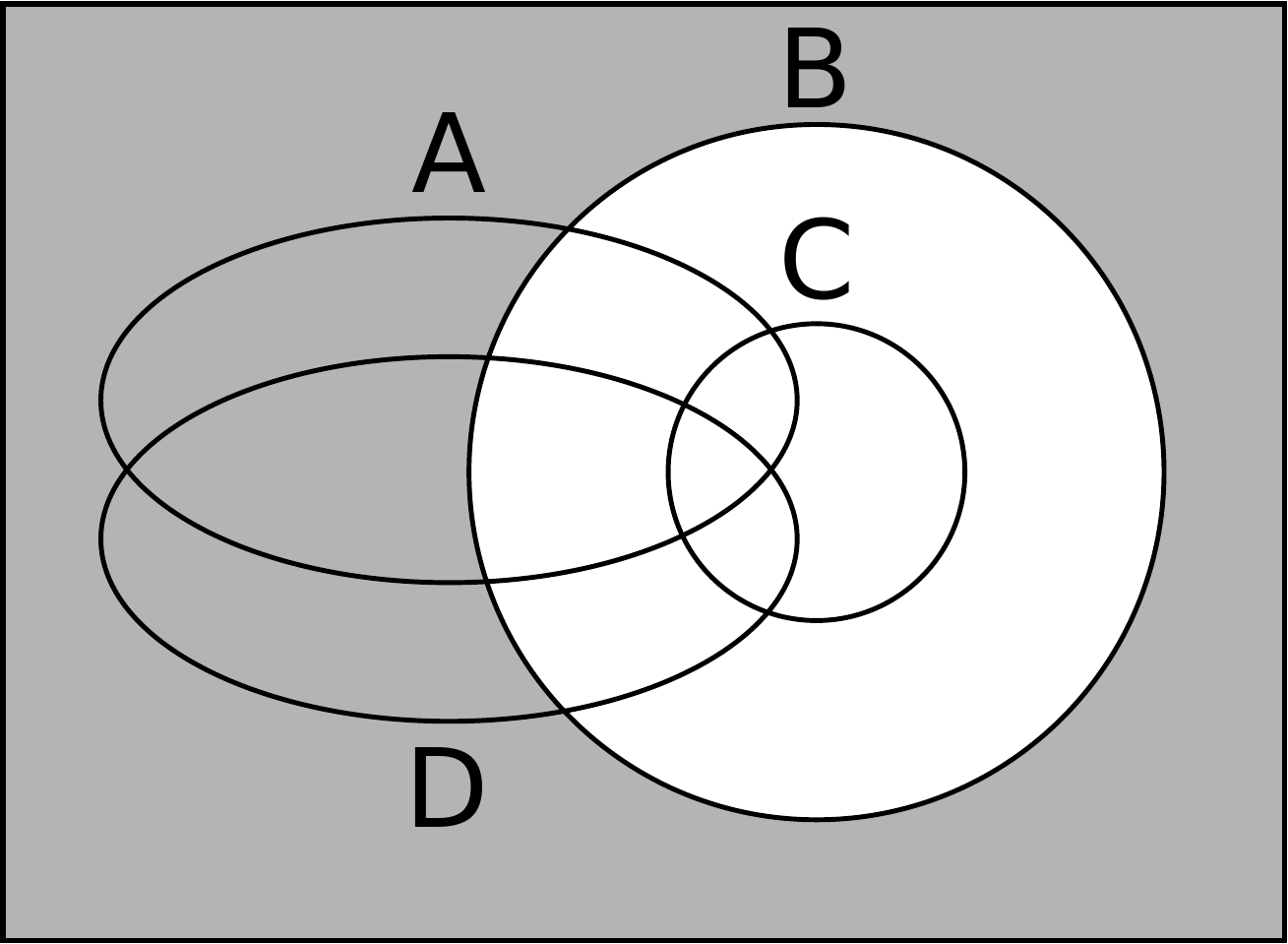}};
     \end{tikzpicture}
   };
   \node[below right =of init, yshift=2.15cm] (tac2) {
     \begin{tikzpicture}
       \node[draw,rectangle](left_conj) {
         \begin{tikzpicture}
           \node (1) {\includegraphics[height=1.5cm]{figures/intCont1}};
           \node[right=of 1, xshift=-1.2cm] (and) {\(\land\)};
           \node[right=of and, xshift=-1.2cm] (2) {\includegraphics[height=1.5cm]{figures/intCont2}};
         \end{tikzpicture}
       };
       \node[right=of left_conj, xshift=-1.05cm] (outer_and) {\(\land\)};
       \node[ right=of outer_and, xshift=-1.05cm](right_conj) {\includegraphics[height=1.5cm]{figures/prem3}};
     \end{tikzpicture}
   };
   \draw[->](init) to node[above] {Tac.~\ref{tac:iac}} (tac1);
   \draw[->](init) to node[below] {Tac.~\ref{tac:iacd}} (tac2);
 \end{tikzpicture}    
 \caption{Examples for Introduce All Contours}
 \label{fig:ex_intr_cont}
\end{figure}

\begin{tac}[Introduce All Contours]
\label{tac:iac}
This tactic (see Figure~\ref{fig:ex_intr_cont}) computes the union, \(C\), of the contour sets in the
antecedent. Then, for each unitary diagram, \(d\), it introduces the
contours from \(C\) that are not present in \(d\). This tactic uses
rule~\ref{r:ic}.
\end{tac}

\begin{tac}[Introduce All Contours (Deepest)]
\label{tac:iacd}
Similarly to the relationship between tactics \ref{tac:iamz} and
\ref{tac:iamzd}, this tactic (see Figure~\ref{fig:ex_intr_cont})
searches for a conjunction of two unitary diagrams and then behaves in
the same way as tactic \ref{tac:iac}. That is, it computes the union
of contour sets and introduces the missing contours into the diagrams.
\end{tac}

\begin{tac}[Combine All Diagrams]
  \label{tac:cmp}
  This tactic searches for conjunctions of unitary diagrams which
  share the same set of zones, and combines each conjunction into a
  single unitary diagram. This is done iteratively; if the result of
  the combination is in conjunction with a suitable diagram, those
  diagrams will also be combined. This tactic uses rule~\ref{r:co}.
\end{tac}


The next two tactics are used within high-level tactics to copy 
elements between conjunctions of unitary diagrams. They provide
a more refined way to introduce and remove zones within diagrams. 

\begin{tac}[Prepare for Copy Shading]
\label{tac:iszfc} 
This specialised tactic introduces new shaded zones into diagrams that are part of a
conjunction, and between which shading can be copied. It uses rule \ref{r:isz}.
\end{tac}

\begin{tac}[Prepare For Copy Contours]
\label{tac:rszfc}
Similar to tactic \ref{tac:iszfc}, this tactic identifies a suitable conjunction and
removes shaded zones from both conjuncts using rule \ref{r:rsz}. 
\end{tac}

The last low-level tactic is intended to be used as the last step
of a derivation, when the  antecedent consists of a single unitary 
diagram that contains all the information needed to reach the consequent. 
This tactic tries to add and remove elements to transform the antecedent into 
the consequent. 

\begin{tac}[Match Conclusion]
\label{tac:mc}
This tactic begins by computing the contour sets,
\(A\) and \(C\), of the antecedent and consequent respectively. It then
introduces each contour \(c \in C \setminus A\) into the diagrams
within the antecedent and
subsequently erases each contour \(c \in A \setminus C\) from these
diagrams. Next, the tactic compares the zones present in the antecedent
and the consequent, and introduces zones that are missing in the
antecedent but present in the consequent. It then erases shading from
zones that are shaded in the antecedent and not shaded in the
consequent. Finally, it transforms all zones that are still shaded in
the antecedent and not present in the consequent into missing zones. To
achieve this, the tactic uses rules \ref{r:ec}, \ref{r:es}, \ref{r:ic},
\ref{r:isz} and \ref{r:rsz}.
\end{tac}

\subsection{High-Level Tactics}
\label{sec:high-level-tactics}

The tactics of this section use one or more rule-level or low-level tactics to create more powerful
tactics. Some of these tactics are powerful enough to solve all subgoals where the conclusion
is a unitary diagram.

The following two tactics copy elements within a conjunction of
unitary diagrams and try to reduce the conjunction to a single unitary
diagram. They can solve a subgoal as long as only contours or shadings
have to be copied.

\begin{tac}[Copy Contours]
\label{tac:cti}
This tactic begins by identifying a conjunction in which a contour can be copied from one
conjunct to the other. It then transforms the shaded zones in this diagram into missing zones to
optimise the effect of the copy process. Subsequently, the tactic copies all contours that can be
copied between the diagrams. Furthermore, if the conjunction consists of two identical
diagrams, it uses idempotency to remove one of the conjuncts.  Hence, it uses rule \ref{r:cc},
idempotency and tactic~\ref{tac:rszfc}.  
\end{tac}

\begin{figure}[htb]
  \centering
  \begin{tikzpicture}
   \node (init) {  
     \begin{tikzpicture}
       \node[draw,rectangle](left_conj) {
         \begin{tikzpicture}
           \node (1) {\includegraphics[height=1.5cm]{figures/prem1}};
           \node[right=of 1, xshift=-1.2cm] (and) {\(\land\)};
           \node[right=of and, xshift=-1.2cm] (2) {\includegraphics[height=1.5cm]{figures/prem2}};
         \end{tikzpicture}
       };
     \end{tikzpicture}
   };
   \node[right  =of init] (tac1) {
     \begin{tikzpicture}
       \node[draw,rectangle](left_conj) {
         \begin{tikzpicture}
           \node (1) {\includegraphics[height=1.5cm]{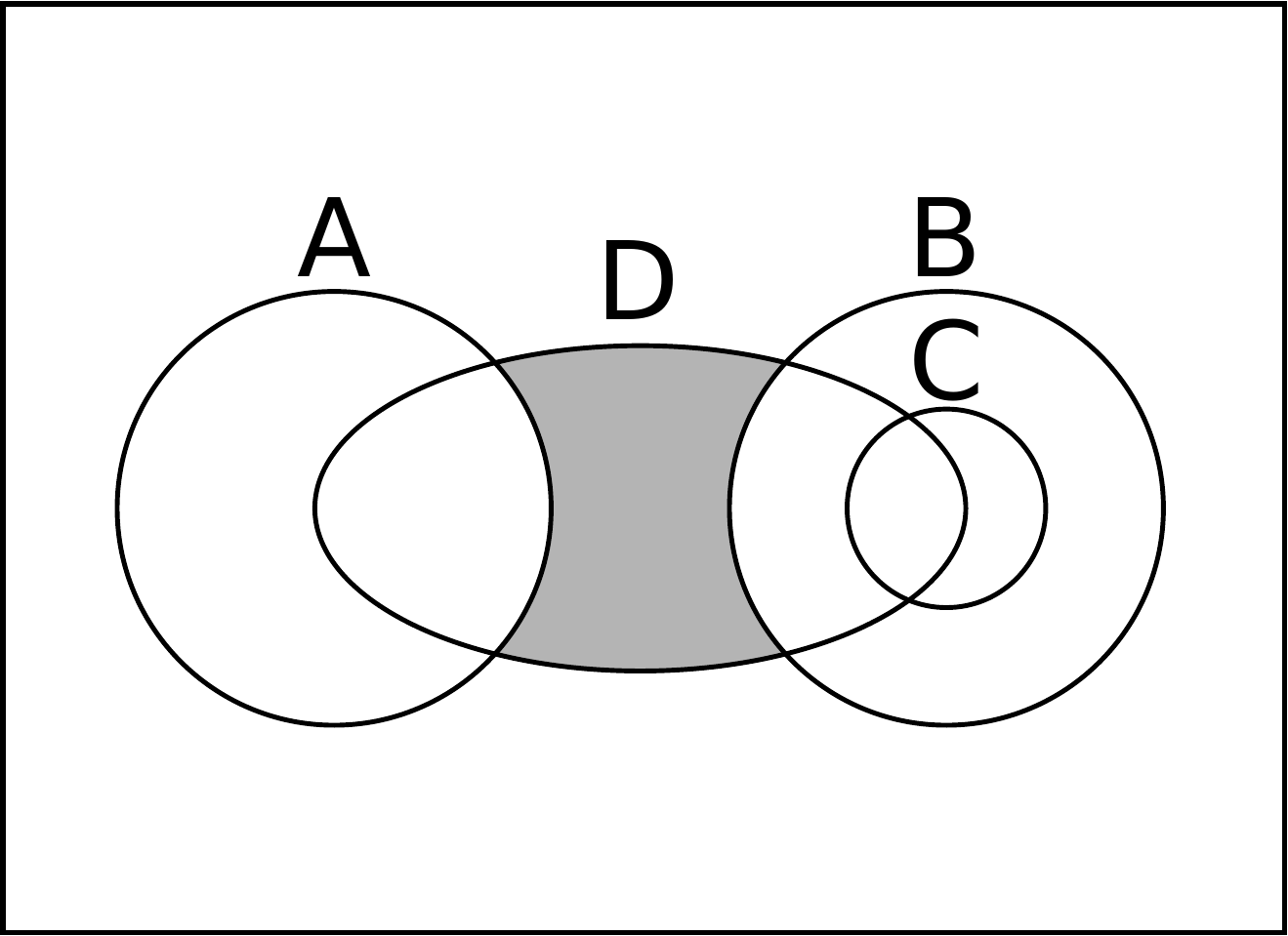}};
           \node[right=of 1, xshift=-1.2cm] (and) {\(\land\)}; 
           \node[right=of and, xshift=-1.2cm] (2) {\includegraphics[height=1.5cm]{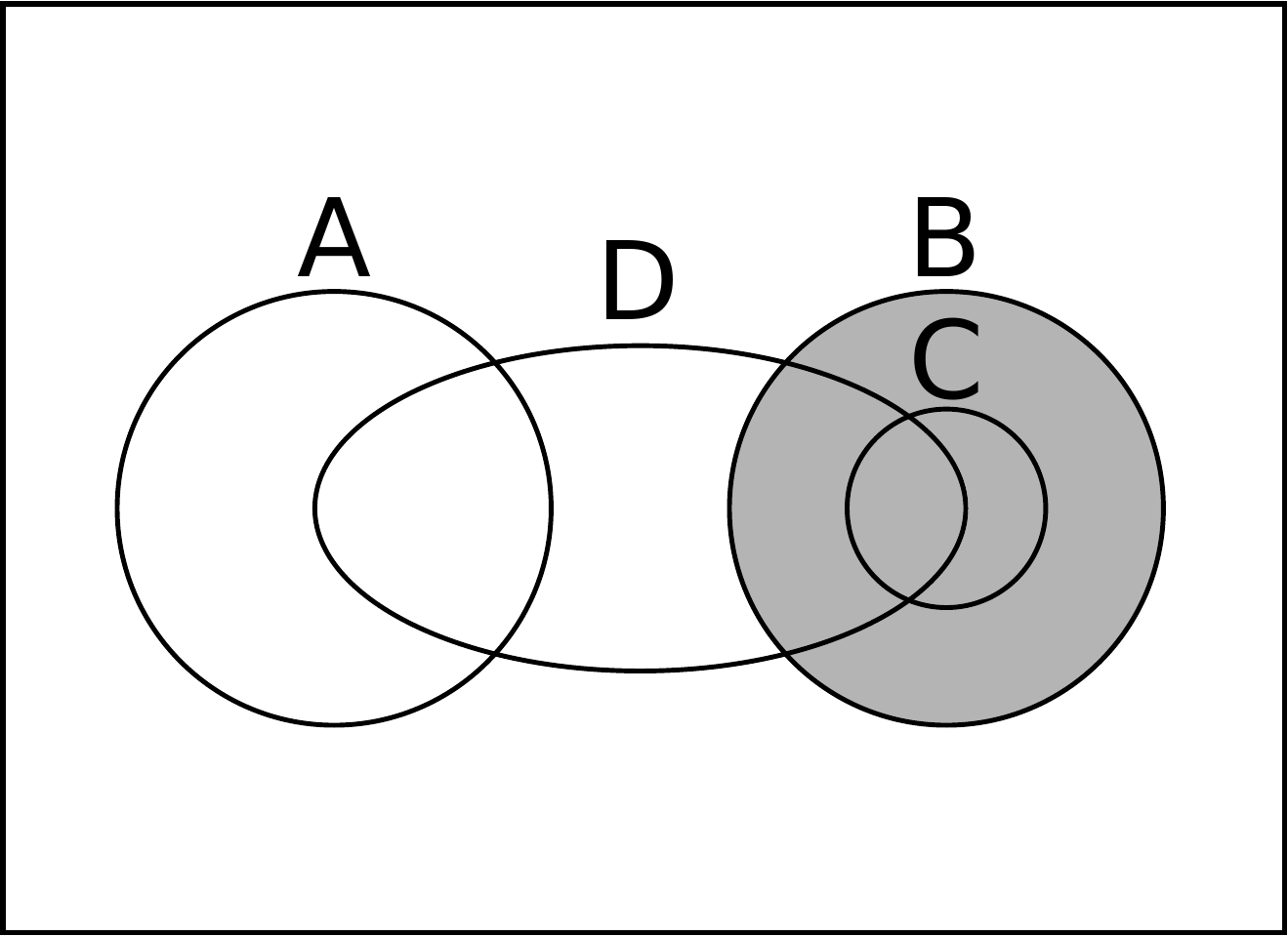}};
         \end{tikzpicture}
       };
     \end{tikzpicture}
   };
   \draw[->](init) to node[above] {Tac.~\ref{tac:cti}} (tac1);
 \end{tikzpicture}
 \caption{Example for Copy Contours}
 \label{fig:ex_copy_contours}
\end{figure}

An example of applying this tactic is shown in Figure~\ref{fig:ex_copy_contours}. The contours
are copied between the two diagrams so that the topological relations between the contours
are preserved. Observe that the shading information is not taken into account and is not
transferred between the diagrams. 

\begin{tac}[Propagate Shading]
\label{tac:csi}
This tactic works similarly to tactic \ref{tac:cti} in that it identifies a suitable
conjunction to copy shading information, then transforms missing zones to shaded zones and ends
by copying the information between the conjuncts. In the conversion phase, it groups the
missing zones according to the contours they include and then introduces one of these groups.
This tactic uses rule \ref{r:cs}, idempotency and tactic~\ref{tac:iszfc}.
\end{tac}

\begin{figure}[htb]
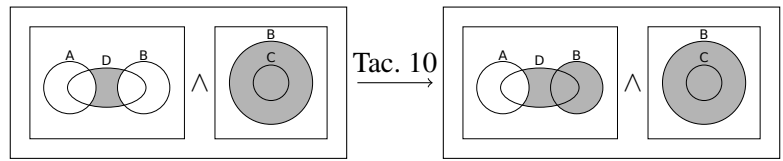

  \centering
  \begin{tikzpicture}
   \node (init) {  
     \begin{tikzpicture}
       \node[draw,rectangle](left_conj) {
         \begin{tikzpicture}
           \node (1) {\includegraphics[height=1.5cm]{figures/prem1}};
           \node[right=of 1, xshift=-1.2cm] (and) {\(\land\)};
           \node[right=of and, xshift=-1.2cm] (2) {\includegraphics[height=1.5cm]{figures/prem2}};
         \end{tikzpicture}
       };
     \end{tikzpicture}
   };
   \node[right  =of init] (tac1) {
     \begin{tikzpicture}
       \node[draw,rectangle](left_conj) {
         \begin{tikzpicture}
           \node (1) {\includegraphics[height=1.5cm]{figures/copyShadingConc}};
           \node[right=of 1, xshift=-1.2cm] (and) {\(\land\)}; 
           \node[right=of and, xshift=-1.2cm] (2) {\includegraphics[height=1.5cm]{figures/prem2}};
         \end{tikzpicture}
       };
     \end{tikzpicture}
   };
   \draw[->](init) to node[above] {Tac.~\ref{tac:csi}} (tac1);
 \end{tikzpicture}
 \caption{Example for Propagate Shading}
 \label{fig:ex_copy_shading}
\end{figure}

Figure~\ref{fig:ex_copy_shading} shows an example of applying \emph{Propagate Shading}. The tactic
does not change the contours present in either diagram. Hence, only the shading of \(B\)
can be transferred. The shading within the left diagram can not be propagated to the right
diagram, since no region corresponds to this shaded zone.

The next two tactics implement ``Venn-style'' reasoning. They transform the diagrams in the
antecedent into Venn-form, combine them until only a unitary diagram remains, and then try to
match the antecedent with the consequent. The difference between the two tactics is the order
in which the steps are performed. In the breadth-first approach, all unitary diagrams in the
antecedent are transformed into Venn-form before any diagrams are combined. Within the
depth-first approach, only a single conjunction of unitary diagrams is transformed into
Venn-form, which is then combined to a unitary diagram, a process which is repeated until only
a single unitary diagram remains.

\begin{figure}[htb]
  \centering
  \begin{tikzpicture}
   \node (init) {  
     \begin{tikzpicture}
       \node[draw,rectangle](left_conj) {
         \begin{tikzpicture}
           \node (1) {\includegraphics[height=1.5cm]{figures/prem1}};
           \node[right=of 1, xshift=-1.2cm] (and) {\(\land\)};
           \node[right=of and, xshift=-1.2cm] (2) {\includegraphics[height=1.5cm]{figures/prem2}};
         \end{tikzpicture}
       };
     \end{tikzpicture}
   };
   \node[above right  =of init, yshift=-2.15cm, xshift=.5cm] (tac1) {
     \begin{tikzpicture}
           \node (1) {\includegraphics[height=1.5cm]{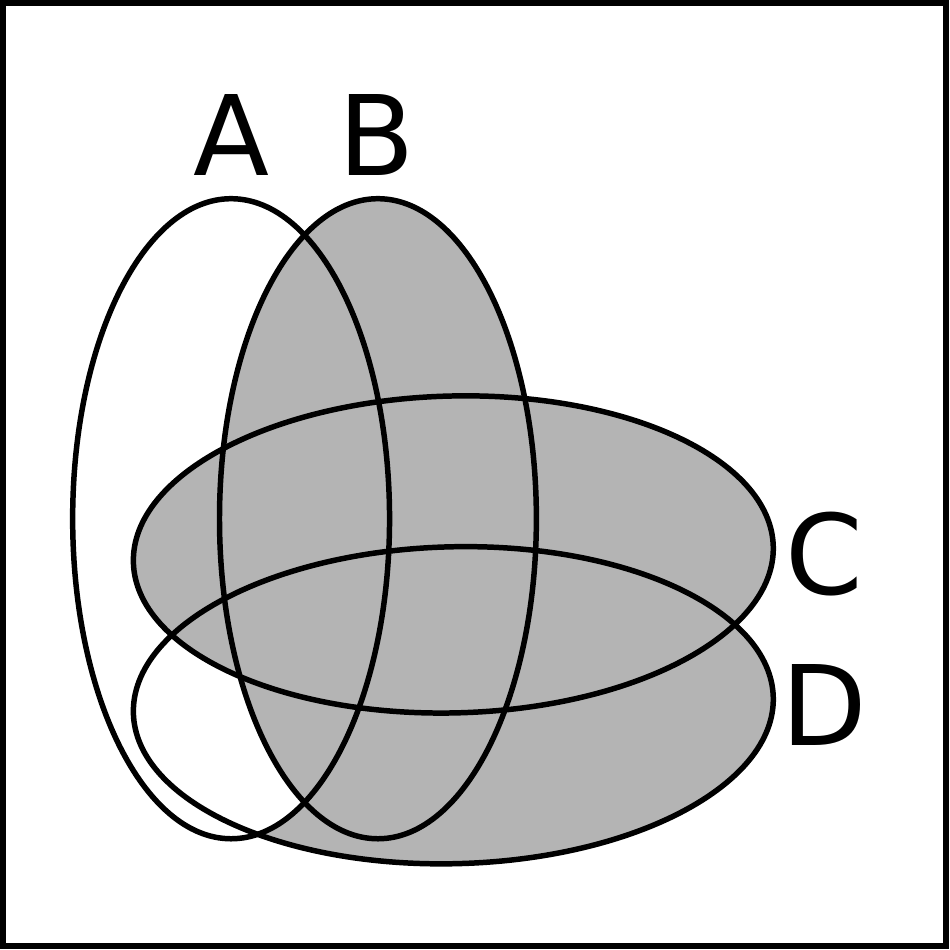}};
     \end{tikzpicture}
   };
   \node[right=of tac1] (match1) {\dots};
   \node[below right =of init, yshift=2.15cm] (tac2) {
     \begin{tikzpicture}
          \node (1) {\includegraphics[height=1.5cm]{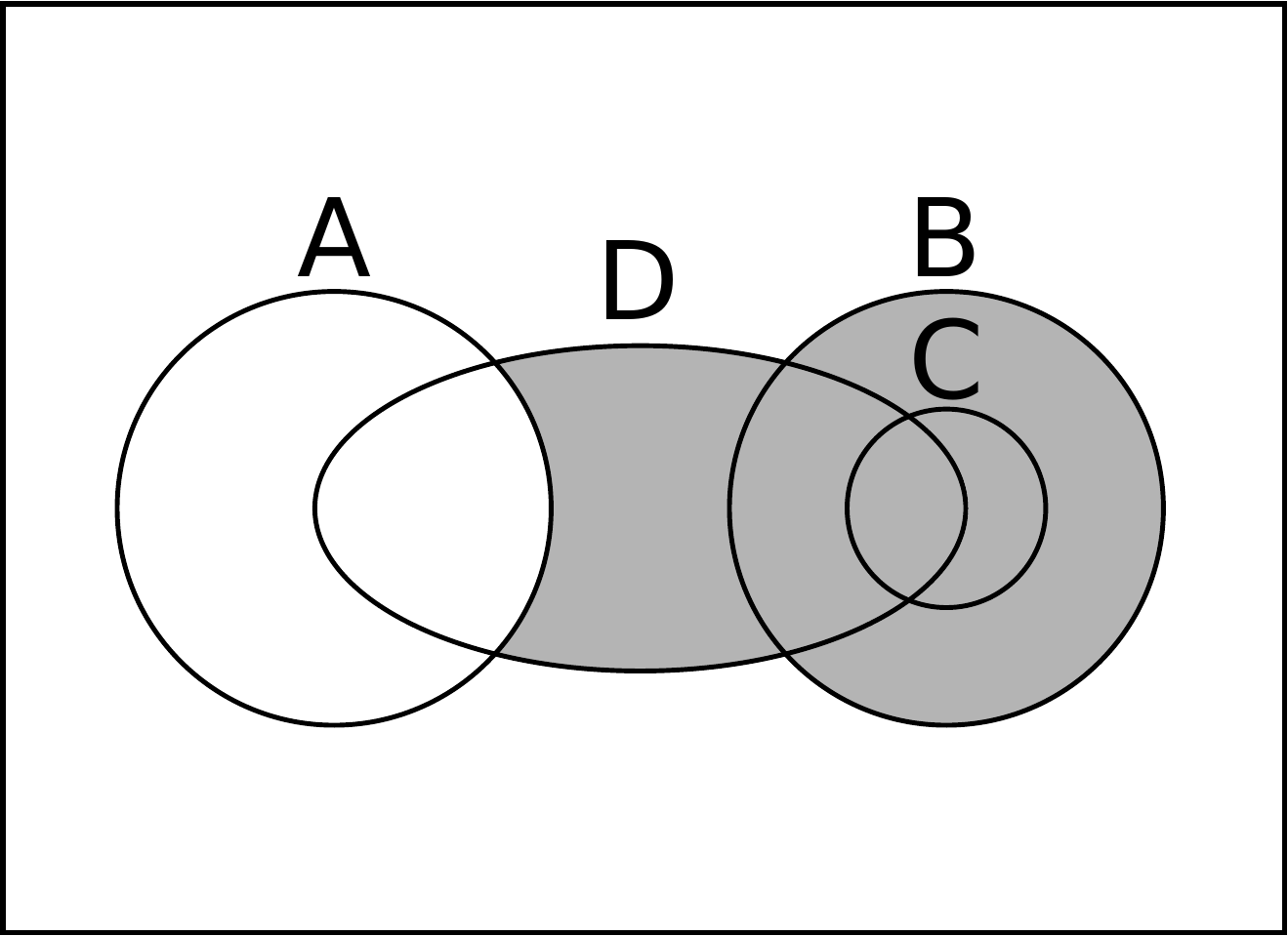}};
       
     \end{tikzpicture}
   };
   \node[right=of tac2] (match2) {\dots};

   \draw[->](init) to node[above] {Tac.~\ref{tac:vb}/\ref{tac:vd}} (tac1);
   \draw[->](init) to node[below] {Tac.~\ref{tac:camap}} (tac2);

   \draw[->](tac1) to node[above] {Tac.~\ref{tac:mc}} (match1);
   \draw[->](tac2) to node[below] {Tac.~\ref{tac:mc}} (match2);
 \end{tikzpicture}    
 \caption{Examples for Venn-Style and Copy Tactic}
 \label{fig:ex_auto}
\end{figure}

\begin{tac}[Venn (Breadth)]
\label{tac:vb}
This tactic (see Figure~\ref{fig:ex_auto}) employs several other tactics to solve a subgoal in the following way. First, it
converts all unitary diagrams in the antecedent into Venn diagrams by introducing missing zones. It then adds contours to
the resulting diagrams until they all contain  the same set of zones, though shading within the
diagrams may differ. 
The tactic then combines conjunctions until only a single unitary diagram
remains using
the \emph{REPEAT} tactical -- note that combining diagrams will fail if only one diagram
remains.
 This diagram is then matched 
with the diagram in the consequent. This tactic uses tactics
\ref{tac:iamz}, \ref{tac:iac}, \ref{tac:cmp} and \ref{tac:mc}.
\end{tac}

\begin{tac}[Venn (Depth)]
\label{tac:vd}
The idea behind this tactic is similar to tactic~\ref{tac:vb}, in that
it also creates Venn diagrams then combines them. However, this tactic
(see Figure~\ref{fig:ex_auto}) chooses and works on one of the
\emph{innermost} conjunctions of unitary diagrams.  When this
conjunction is combined to a single unitary diagram, the process
starts from the beginning by selecting a new innermost conjunction,
until only one unitary diagram remains, using \emph{DEPTH\_FIRST}
search.  Finally, this diagram is matched against the consequent. This
tactic uses tactics \ref{tac:iamzd}, \ref{tac:iacd}, \ref{tac:cmp}
and \ref{tac:mc}.
\end{tac}

The last high-level tactic combines the tactics for copying
elements. It applies both copying tactics~\ref{tac:cti}
and~\ref{tac:csi} until it reaches the conclusion.

\begin{tac}[Copy Shading And Contours]
\label{tac:camap}
This tactic (see Figure~\ref{fig:ex_auto}) uses the copy tactics to
collect all information from the antecedent within one unitary
diagram. It applies tactic \ref{tac:csi} and then tactic \ref{tac:cti}
until only a unitary diagram remains (note that both of these tactics
seek to reduce the number of unitary diagrams in the antecedent by
using idempotency). Again, this tactic uses \emph{DEPTH\_FIRST} to
check when the result is unitary. The resulting diagram is matched to
the consequent using tactic \ref{tac:mc}.
\end{tac}

Figure~\ref{fig:ex_auto} shows the state during the execution
of either of the Venn-style tactics and of the Copy tactic, right before the 
tactics try to match the antecedent to the consequent. The Venn-style
tactics create a Venn-diagram, where all the information of the antecedent
is visible through the shading. In contrast, in the case of the Copy tactic,
part of this information is visualised through the topological relations
between the contours. Both diagrams contain the same semantic information.

Of course, these are not the only possible high-level tactics one could define. We chose this
set of tactics for the following reasons. Venn-style reasoning has been traditionally used for
Euler-based diagram languages. For example, in the original definition of Spider diagrams
\cite{howse:sd}, the only rule usable for Euler diagrams to transfer information from one
diagram to the other is \emph{Combine}, which requires diagrams with the same set of zones as
input. Hence Venn-style reasoning is the only way to join the information contained within
several diagrams. However, Speedith \cite{Urbas2015} allows for the use of the
copy rules, and we have found the copying tactics provide effective reasoning strategies in
many situations. Furthermore, we do not allow for user-defined tactics. That is, in order to
create new tactics in addition to the ones described above, users would have to write Scala
code directly. This is necessary because writing a new tactic requires knowledge of the
implementation of the individual rules and tactics on which the new tactic would depend.






\section{\label{sec:eval}Evaluation}

The previous version of Speedith functioned as a simple proof
assistant, enabling the application of individual rules. In our
extension, the tool can still be used in the traditional way, and our
tactical reasoning features appear in a separate menu. We also added a
number of more minor enhancements related to usability, such as the
ability to save and load proofs, and an undo capability which allows
the user to return to any subgoal by discarding the subsequent steps. 

We carried out an empirical study on the effects of clutter on readability of
diagrammatic proofs, which showed that participants took significantly longer
to identify rule applications in cluttered
diagrams~\cite{linker:mucoiried}. Thus, we use various metrics about
the readability of proofs to guide the choices of tactics used in
proofs.  In our version of the tool, we calculated the following
metrics for proofs: the length of the proof, the total and average
clutter of the proof (where clutter is calculated as the number of all
zones present plus the number of shaded zones) and the maximum
\emph{clutter velocity} (the largest change in clutter between one
subgoal and its subsequent subgoal). These metrics are available
through the interface and we used them to compare alternative proofs and
tactics.

\begin{figure}[t]
  \begin{subfigure}{\linewidth}
    \begin{center}
      \begin{tikzpicture}
        \node[draw,rectangle] (premiss) {
          \begin{tikzpicture}
            \node[draw,rectangle](left_conj) {
              \begin{tikzpicture}
                \node (1) {\includegraphics[height=2cm]{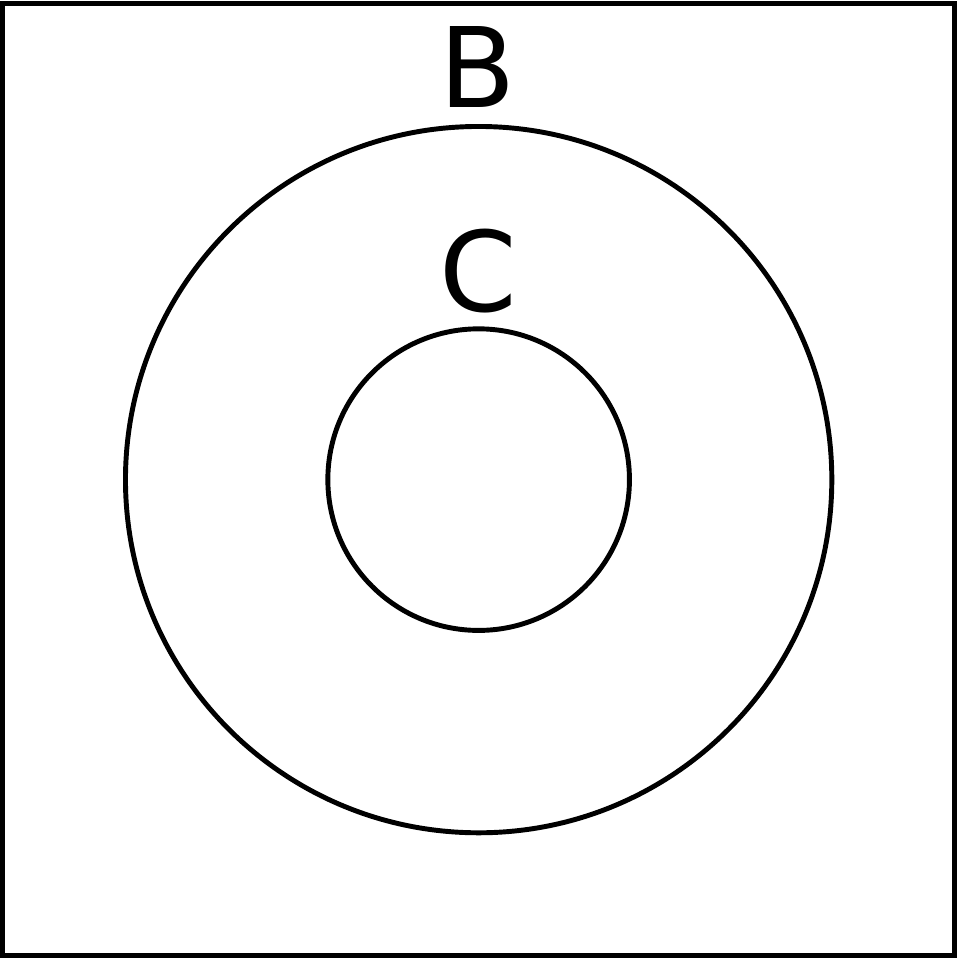}};
                \node[right=of 1, xshift=-1.2cm] (and) {\(\land\)};
                \node[right=of and, xshift=-1.2cm] (2) {\includegraphics[height=2cm]{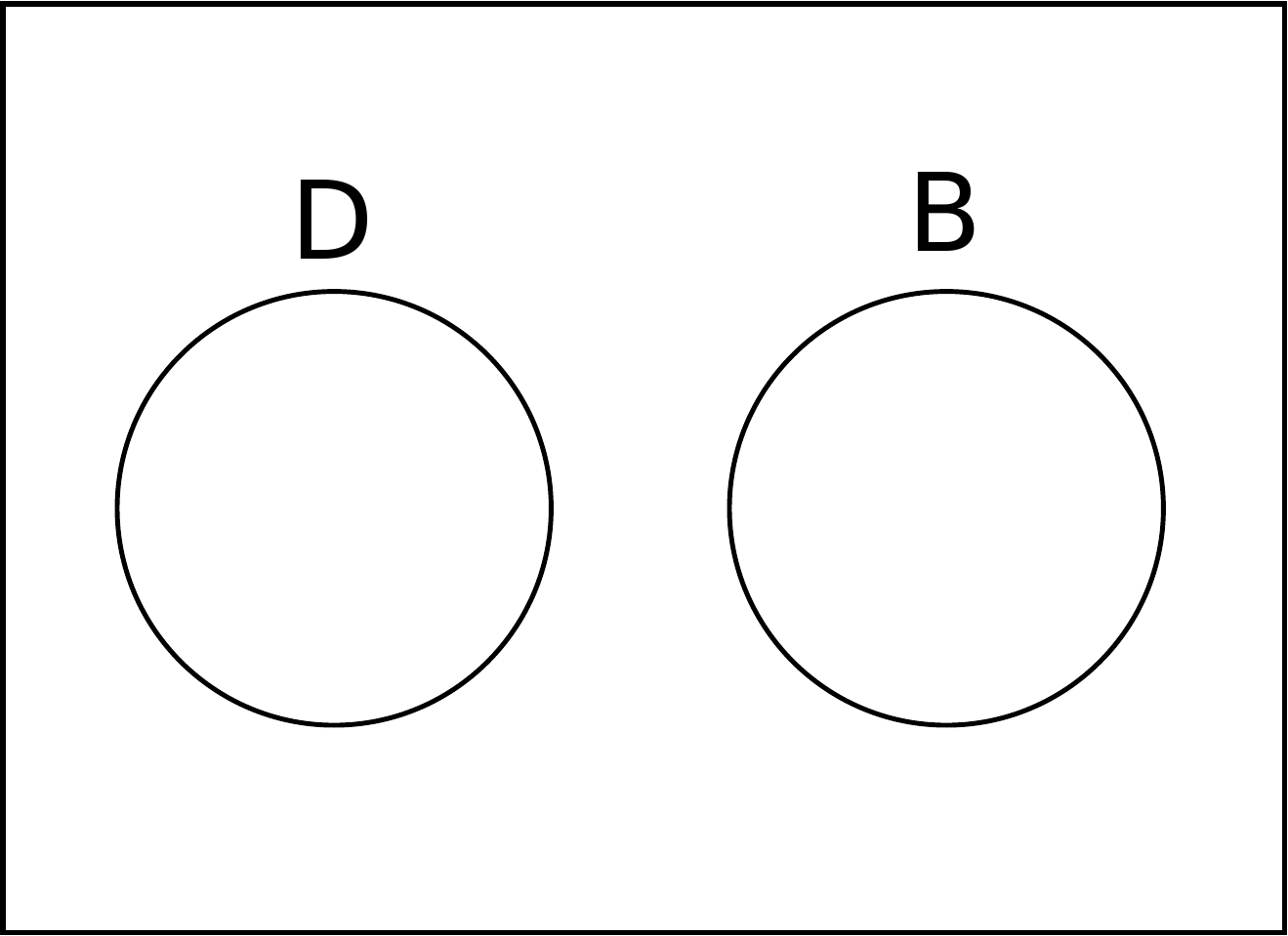}};
              \end{tikzpicture}
            };
            \node[right=of left_conj, xshift=-1.05cm] (outer_and) {\(\land\)};
            \node[draw,rectangle, right=of outer_and, xshift=-1.05cm](right_conj) {
              \begin{tikzpicture}
                \node (1) {\includegraphics[height=2cm]{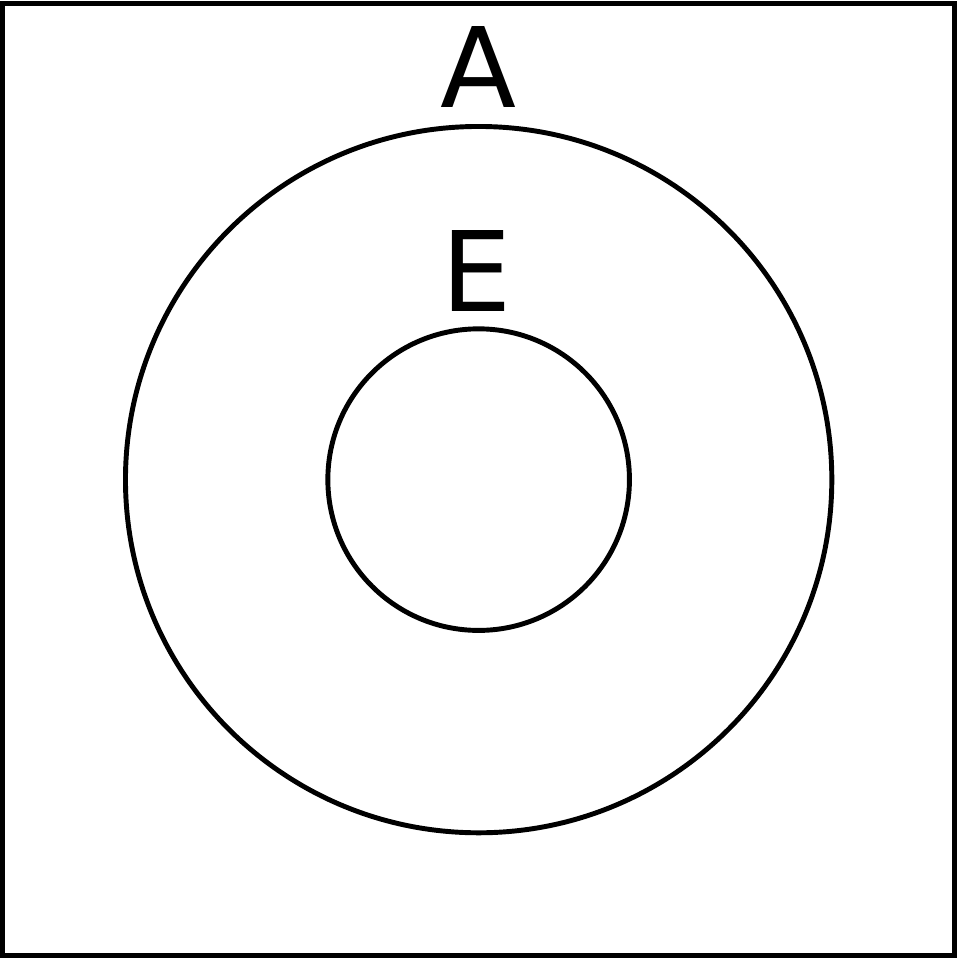}};
                \node[right=of 1, xshift=-1.2cm] (and) {\(\land\)};
                \node[right=of and, xshift=-1.2cm] (2) {\includegraphics[height=2cm]{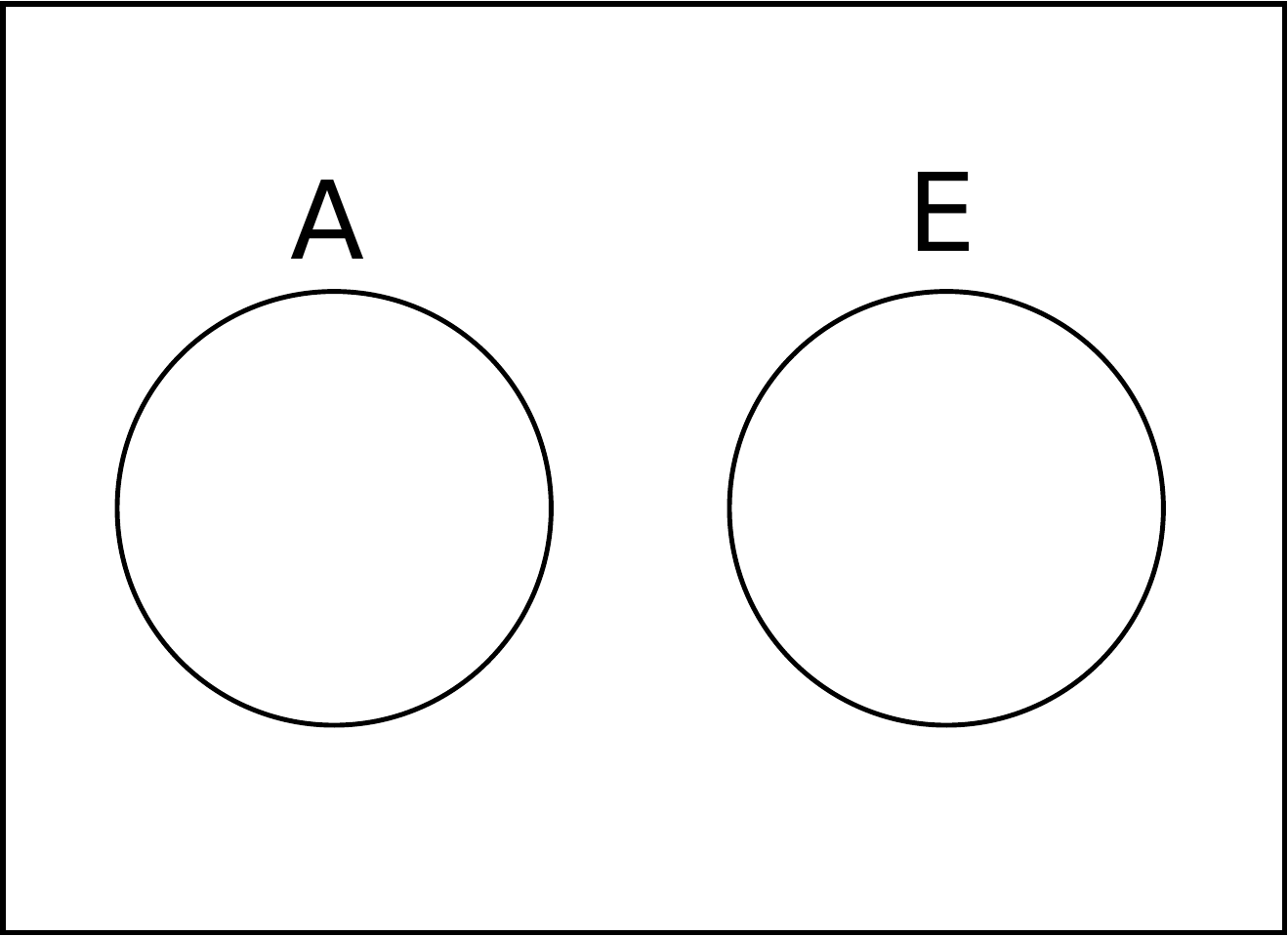}};
              \end{tikzpicture}
            };
          \end{tikzpicture}
        };
        \node[right=of premiss, xshift=-1.05cm] (imp) {\(\imp\)};
        \node[right=of imp, xshift=-1.05cm] (conc) {\includegraphics[height=2.75cm]{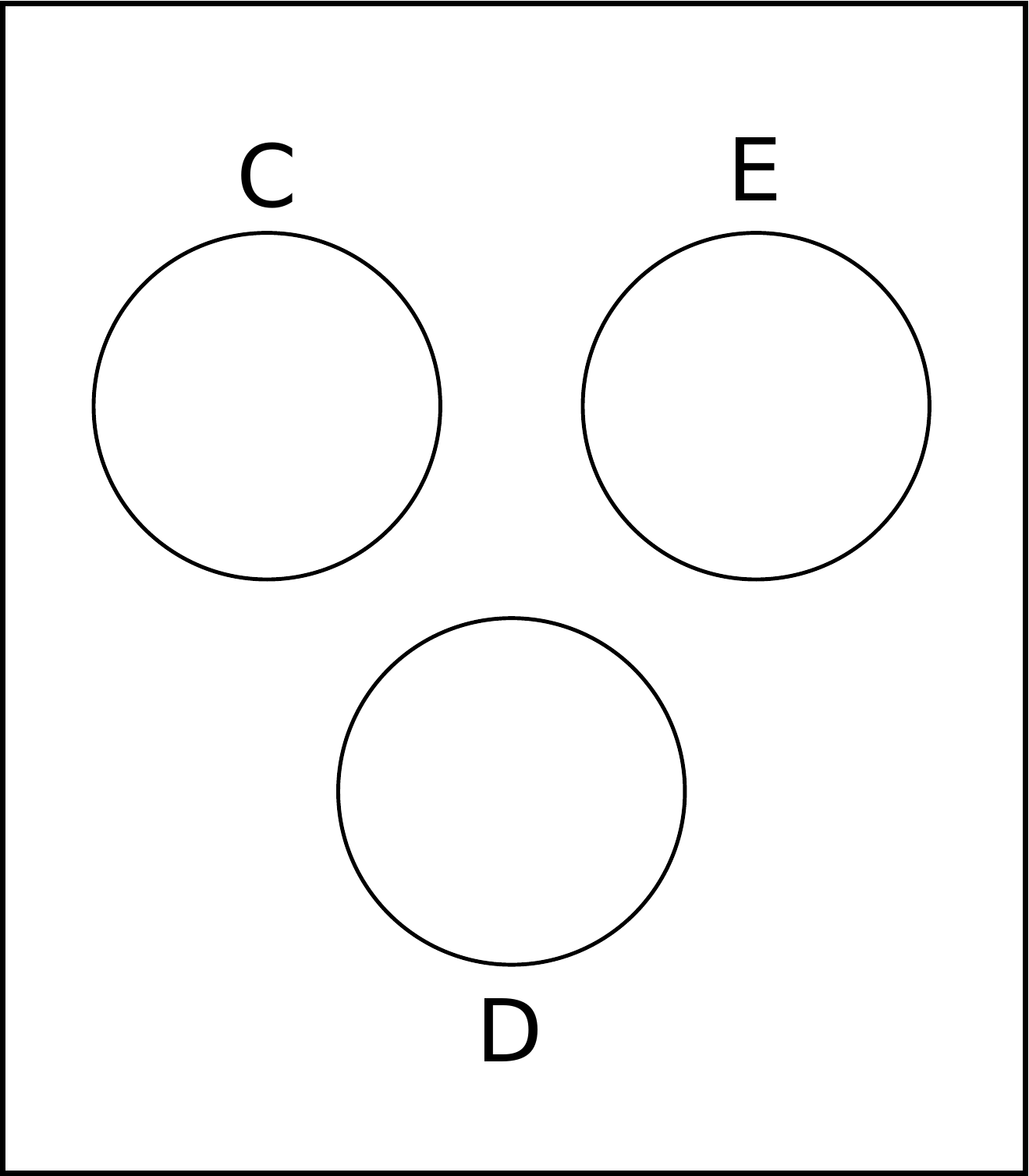}};
      \end{tikzpicture}
      \caption{Flat Conjunctive Structure}
      \label{fig:struc_diff_1}
    \end{center}
  \end{subfigure}

  \begin{subfigure}{\linewidth}
    \begin{center}
      \begin{tikzpicture}
        \node[draw,rectangle] (premiss) {
          \begin{tikzpicture}
            \node[draw,rectangle](outer_conj) {
              \begin{tikzpicture}
                \node (left_conj) {\includegraphics[height=2cm]{figures/c_subset_b}};
                \node[right=of left_conj, xshift=-1.2cm] (outer_and) {\(\land\)};
                \node[draw,rectangle, right= of outer_and,  xshift=-1.05cm](right_conj) {
                  \begin{tikzpicture}
                    \node (1) {\includegraphics[height=2cm]{figures/e_subset_a}};
                    \node[right=of 1, xshift=-1.2cm] (and) {\(\land\)};
                    \node[right=of and, xshift=-1.2cm] (2) {\includegraphics[height=2cm]{figures/d_disjoint_b}};
                  \end{tikzpicture}
                };
              \end{tikzpicture}
            };
            \node[right=of outer_conj, xshift=-1.05cm] (and2) {\(\land\)};
            \node[right=of and2, xshift=-1.2cm] (2) {\includegraphics[height=2cm]{figures/a_disjoint_e}};
          \end{tikzpicture}
        };
        \node[right=of premiss, xshift=-1.05cm] (imp) {\(\imp\)};
        \node[right=of imp, xshift=-1.05cm] (conc) {\includegraphics[height=2.75cm]{figures/c_disjoint_d_disjoint_e}};
      \end{tikzpicture}
      \caption{Deep Conjunctive Structure}
      \label{fig:struc_diff_2}
    \end{center}
  \end{subfigure}
  \caption{Structural Differences in Theorems}
  \label{fig:struc_diff}
\end{figure}

In sentential tactical reasoning it is normally the case that powerful
auto-style tactics are best used when the main concern is that the
reasoner finds a proof, whether or not the user
wants to inspect or understand that proof. Our tactics \ref{tac:vb} and \ref{tac:vd}, for
Venn-style reasoning, can be considered auto-style tactics, and will
frequently reach a proof without further input from the user. However,
an important difference is that tactic \ref{tac:vd} is designed to
work in a way which is perhaps closer to the way a human might attempt
to apply the same strategy. The approach taken by the breadth first
version (tactic \ref{tac:vb}) is effective from a technical point of
view but produces relatively long proofs containing cluttered
diagrams.  Arguably, a more likely human approach would be to maintain
focus on one or two unitary diagrams at a time: add contours and
shaded zones to two unitary diagrams until they have the same zone
set, then combine, then pick another two diagrams to work on, and so
on.  
Our design of tactic~\ref{tac:vd}, which maintains focus on two diagrams at a time, relates to
the ``story telling'' perspective of making and reading proofs, and to their communicative
rather than their formal function. Whether this tactic really does match the approach taken by
humans remains to be proven empirically. However, in many cases the depth first approach also
produces sequences of inferences which are shorter and which contain less cluttered
diagrams. Thus, relative to tactic \ref{tac:vb}, tactic \ref{tac:vd} produces proofs which, according to the results of our
empirical study, will frequently be easier for readers to understand. Although we don't have
room to display entire proofs in this work\footnote{As stated earlier, a body of proofs is available from our
  website at \url{http://readableproofs.org/speedith}}, we measured the performance of the different
approaches when
applied to the same theorem. The theorem shown in Figure~\ref{fig:struc_diff_1} can be proved
by either of the Venn-style tactics, and the metrics for each proof is shown in
Table~\ref{tab:metrics}, where overall clutter is the clutter of the entire proof, showing
better performance for the depth first approach.

%


The behaviour of tactics is heavily reliant on the concrete structure of the theorem to
prove. Consider the examples in Figure~\ref{fig:struc_diff_1} and
\ref{fig:struc_diff_2}. Table~\ref{tab:metrics2} shows the metrics of proofs obtained by using
tactic~\ref{tac:camap} (Copy Shading and Contours) to each diagram, where maximal clutter is
the sum of the clutter in the most cluttered proof step. Here we see a marked difference in the
metrics, and our study predicts that the proof of the theorem in Figure~\ref{fig:struc_diff_1} would be
significantly easier for a user to understand, 
relative to the proof of the equivalent theorem
in Figure~\ref{fig:struc_diff_2}. This difference is mostly
due to the fact that in Figure~\ref{fig:struc_diff_1} each of the inner conjunctions allows for
the application of exactly one of the copy tactics: the left conjunct only allows for the
application of \emph{Copy Contours} (tactic \ref{tac:cti}), while the right conjunct can only
be reduced by applying \emph{Propagate Shading} (tactic \ref{tac:csi}). Subsequently, the
remaining contours can be copied between the resulting diagrams. For
Figure~\ref{fig:struc_diff_2}, however, all contours need to be copied before anything else is
done, so that the resulting conjunction contains all 5 contours. Finally, tactic~\ref{tac:csi} can be applied to this conjunction, which introduces zones and thus increases the
clutter.

{\centering
\begin{table}[ht]
\begin{minipage}[b]{0.45\linewidth}
\centering

\caption{Two proofs of the theorem in Figures~\ref{fig:struc_diff_1}}\label{tab:metrics}
\begin{tabular}[t]{|l|l|l|}
\hline
Tactic & Steps & Overall clutter \\
\hline
Venn (Breadth) & 27 & 355 \\
Venn (Depth)   & 22 & 195 \\

\hline
\end{tabular}
\end{minipage}
\hspace{0.5cm}
\begin{minipage}[b]{0.45\linewidth}
\centering

\caption{Applying Tactic~\ref{tac:camap} to the theorems in Figs~\ref{fig:struc_diff_1} and~\ref{fig:struc_diff_2}}\label{tab:metrics2}
\begin{tabular}[t]{|l|l|l|}
\hline
Theorem & Steps & Maximal clutter \\
\hline
Figure~\ref{fig:struc_diff_1} & 20 & 27 \\
Figure~\ref{fig:struc_diff_2} & 28 & 201 \\

\hline
\end{tabular}

\end{minipage}
\end{table}
} 

 


\section{Conclusion}

By adding tactical reasoning to Speedith, we produced the first diagrammatic reasoning system
with this powerful form of semi-automation. We did so in a way that seeks to exploit the
distinctively diagrammatic features of the logic, and which takes advantage of what is known
about the cognitive benefits and shortcomings of Euler-based notations. We aimed to produce a
system that can aid the user to automatically produce proofs which model an approach to problem
solving we expect to be more coherent to a human reader and which exhibit features, such as
containing relatively less clutter, which we have empirically established as having an positive
impact on understanding. Our analysis of tactics and proofs using the metrics described in the
previous section enables us to differentiate between several approaches to proving a particular
theorem in quantitative ways.

Even though our implementation was inspired by Isabelle, our tactic combinators are typical for systems combining rules
in a (semi-) automated fashion. For example, the sequential composition \emph{THEN}
and the repeated application \emph{REPEAT} and \emph{DEPTH\_FIRST} are similar to 
constructs of Quantomatic \cite{Kissinger2015}.  Our combinators  
only create  single  proof states, that is, it is not possible to use backtracking if applying a 
tactic fails. Extending the implementation to return lists of possible
proof states is future work. Furthermore, we intend to implement rules that
split conjunctions and disjunctions in appropriate positions into several subgoals. Then, the need for new
 tactics to split and reduce subgoals naturally arises.  

One of our main goals is pedagogical, and we aim to use Speedith and its tactics to teach
diagrammatic reasoning. We expect that this type of learning will be supported by allowing the
user to experiment with diagrams more directly than is currently possible in Speedith; that is,
by drawing, manipulating and experimenting with diagrams themselves. To support this goal, we
are working on a Speedith plugin for Openproof, in which Speedith's core reasoning engine and inference
rules are made available alongside an Euler diagram editor. Diagrams drawn by the user are
converted to the abstractions used by Speedith and checked for validity. In this way, we expect
to provide an accessible and novel tool for teachers and students of logic. Finally, although
the graphical library we use, iCircles, can draw every Euler diagram, it sometimes produces
diagrams with undesirable features such as duplicate labels (whereby a set is represented by
two disjoint circles with the same label), and the layout of its diagrams may be suboptimal in
other ways. In order to address this, work is underway to improve the layout algorithm of
iCircles.



\bibliographystyle{eptcs}
\bibliography{uitp}
\end{document}